\documentclass
[twocolumn]
{jpsj3}
\usepackage{multirow}
\usepackage{bm}
\makeatletter

\@addtoreset{equation}{section}
\makeatother

\title{Kondo Effect of a Magnetic Ion Vibrating in a Harmonic Potential}
\author{Satoshi YASHIKI\thanks{E-mail address: syashiki@issp.u-tokyo.ac.jp}, Shunsuke KIRINO, Kazumasa HATTORI
        and Kazuo UEDA}
\inst{Institute for Solid State Physics, University of Tokyo, Kashiwa 277-8581, Japan}
\abst{
To discuss Kondo effects of a magnetic ion vibrating in the sea of conduction electrons, a generalized Anderson model 
is derived. The model includes a new channel of hybridization associated with phonon emission or absorption. In the 
simplest case of the localized electron orbital with the $s$-wave symmetry, hybridization with $p$-waves becomes possible. 
Interesting interplay among the conventional $s$-wave Kondo effect and the $p$-wave one and the Yu-Anderson type Kondo effect 
is found and the ground state phase diagram is determined by using the numerical renormalization group method. 
Two different types of stable fixed points are identified and the two-channel Kondo fixed points are generically realized 
on the boundary.
}
\kword{Kondo effect, impurity Anderson model, electron-phonon coupling, numerical renormalization group, 
two-channel Kondo effect}

\begin{document}
\maketitle
\section{Introduction}
Recently, a new type of ionic structure, a network of cages filled or unfilled by guest ions, has attracted much
interest in the research area of condensed matter physics. When the radius of filled ion is smaller than 
the size of cage, the guest ions vibrate with large amplitude but almost without effect of interference with oscillations 
of the ions in the neighboring cages. It is expected that various unusual phenomena are possible due to such incoherent vibrations. 
Typical examples of this class of materials are $\beta$-pyrochlore oxides AOs$_{2}$O$_{6}$(A $=$ K, Rb or Cs), filled 
skutterudites RT$_{4}$X$_{12}($R $=$ rare earth or alkaline earth; T $=$ Fe, Ru, Pt or Os; X $=$ P, As, Ge or Sb$)$ and 
type-I clathrate compounds.

The family of $\beta$-pyrochlore oxides, especially KOs$_{2}$O$_{6}$, is a good example where several qualitatively 
new phenomena are observed due to the anharmonicity of the cage {potential\cite{Nagao}}: concave-downward 
temperature dependence of 
{resistivity\cite{Hiroi}}, peak structure of NMR relaxation rate at potassium site at low {temperatures\cite{Yoshida}}, 
occurrence of superconductivity\cite{Hiroi2} and the structural phase transition which takes place below the 
superconducting transition temperature\cite{Hiroi}. 
Concerning the resistivity and the NMR relaxation rate, 
Dahm and one of the authors considered a simple model for the anharmonic potential which includes the fourth order term 
of the ion displacement. They have shown that the unusual temperature dependence of these quantities are explained by calculating 
the retarded phonon propagator within the self-consistent quasi-harmonic approximation\cite{Dahm_Ueda}. Detailed 
properties of the spectral function of the local ion are also investigated for various types of the cage potential including 
a double-well type\cite{Takechi_Ueda}. The more realistic model that the ion vibrates in the three-dimensional cage 
potential with the third order anharmonic term specific to the tetrahedral symmetry of the $\beta$-pyrochlore compounds 
is discussed by one of the authors and Tsunetsugu\cite{Hattori1,Hattori2,Hattori4}. It is shown that the third order anharmonic 
term has important effect on the superconductivity and is also responsible to the first order structural 
transition below the superconducting transition temperature.

Another interesting behavior is reported for the filled skutterudite compound, SmOs$_{4}$Sb$_{12}$\cite{Sm_Os4_Sb12_1,
Sm_Os4_Sb12_2,Sm_Os4_Sb12_3,Sm_Os4_Sb12_4,Sm_Os4_Sb12_5}. It shows a large specific heat coefficient $\gamma$, 
which is robust against magnetic fields\cite{Sm_Os4_Sb12_2}. One possible scenario of this behavior is that a non-magnetic 
Kondo effect is realized by local vibrations of the guest ions\cite{Hattori3,Mitsumoto}. Motivated by the unusual heavy 
fermion behaviors of SmOs$_{4}$Sb$_{12}$, we have started a systematic study on the interplay between local electron correlation 
and electron-phonon coupling\cite{Letter}. 

Let us review the previous studies related to Kondo effect coupled with local vibrations. To include effects of 
the local oscillations, two different types of models have been considered, the breathing mode of the cage and the transverse modes of 
the cage, i.e., the relative displacement between the ion and the cage.

Firstly, a theoretical model of a magnetic ion coupled with the breathing mode is known as Anderson-Holstein model\cite{AH_model}. 
In this model the one-electron energy level of the impurity orbital is modified by the amplitude of the breathing mode. 
By applying Lang-Firsov {transformation\cite{Lang-Firsov}}, it can be shown that the interaction induces an effectively 
attractive electron-electron interaction (for example see Ref. \ref{Hotta}). If the electron-phonon coupling overcomes
Coulomb repulsion, the electronic states of the impurity result in a quasi degenerate doublet, the empty and the doubly 
occupied states. If there is a matrix element between the doublet, it is possible that the non-magnetic doublet works 
as a pseudo spin and induces Kondo effect. Actually, a recent study shows that magnetically robust Kondo effect is 
possible even in relatively high magnetic fields when the effect of anharmonicity of the breathing mode is included\cite{Hotta_2}.

Secondly, a model of a non-magnetic ion which oscillates in the sea of conduction electrons was proposed by Yu and 
Anderson\cite{Yu_Anderson}. In this model, the conduction electrons are scattered by the vibrations of the ion. For simplicity 
spins of conduction electrons are neglected. Yu and Anderson considered only the first order term of ionic displacement 
which produced the scattering processes between the $s$-wave conduction electrons to the $p$-wave ones. 
It is shown that the resultant effective potential for the ion displacement behaves like a double well potential when the 
electron-phonon coupling is strong. Therefore a new type of Kondo effect is expected because logarithmic divergence appears 
when there are doubly degenerate states connected each other through scattering processes of conduction 
electrons\cite{TLM1,TLM2,TLM3}.

This paper is a comprehensive report on the effects of vibration of a magnetic ion coupled with spinful conduction electrons.
Some of the results have been published in the previous letter\cite{Letter}.
In \S\ref{section_2} we construct a generalized impurity Anderson model of the magnetic ion.
When we ignore spin-orbit interaction, the generalized gradient theorem\cite{Bayman} enables us to expand hybridization 
term with respect to the ion displacement. As a result phonon-assisted hybridization is naturally induced in addition to the 
conventional one. In \S\ref{section_3} and \S\ref{section_4}, we apply the numerical renormalization group (NRG) 
method\cite{Wilson,krishna} to the simplest case, where the impurity ion with an $s$-wave orbital vibrates in a one-dimensional 
harmonic potential. From the analysis of energy spectra obtained from the NRG calculations, one can identify three different 
types of NRG flows at low temperatures which define the low energy fixed points. We obtain the phase diagram characterized 
by the fixed points in the parameter space of phonon-assisted hybridization and Coulomb interaction.
The fixed point analysis is described in detail in the present paper. Various physical quantities which include the spin 
and charge susceptibilities and the mean square amplitude of the displacement are also discussed. Finally,
conclusions and some future problems are summarized in \S\ref{section_5}.

\section{A generalized impurity Anderson model for a vibrating magnetic ion}\label{section_2}
\subsection{Expansion of hybridization term by ion displacement}
In this section we derive a generalized impurity Anderson model for a magnetic ion vibrating in a cage potential. When the ion 
is static, it has to be equivalent to the usual impurity Anderson model\cite{Anderson_model}. It is assumed that the effects 
of vibrations are taken into account through hybridization modified by the shift of the ionic position. Conduction electrons 
are assumed to be a non-interacting system and decomposed into the partial waves,
\begin{align}
  \mathit{H}_{\text{c}}
&=\sum_{\boldsymbol{k}\sigma}\varepsilon(\boldsymbol{k})c^{\dagger}_{\boldsymbol{k}\sigma}c_{\boldsymbol{k}\sigma}\notag\\
&=\sum_{k\sigma}\sum_{\ell m}\varepsilon(k)c^{\dagger}_{\ell m\sigma}(k)c_{\ell m\sigma}(k),
\end{align}
where $\varepsilon(\boldsymbol{k})$ is the dispersion relation and $c_{\boldsymbol{k}\sigma}$ $(c^{\dagger}_{\boldsymbol{k}\sigma})$ 
is an annihilation (creation) operator of the conduction electrons. $c_{\ell m\sigma}(k)$ $(c^{\dagger}_{\ell m\sigma}(k))$ 
is defined from $c_{\boldsymbol{k}\sigma}$ $(c^{\dagger}_{\boldsymbol{k}\sigma})$ by using the spherical harmonics 
$Y_{\ell m}(\hat{\boldsymbol{k}})$,
\begin{align}
c_{\ell m\sigma}(k)=\sqrt{4\pi}\sum_{\hat{\boldsymbol{k}}}Y^{*}_{\ell m}(\hat{\boldsymbol{k}})c_{\boldsymbol{k}\sigma}.
\end{align}
The dynamics of the ion with position vector $\boldsymbol{Q}$ is described by the following Hamiltonian,
\begin{align} 
\mathit{H}_{\text{ion}}&=-\frac{\hbar^{2}}{2M}(\nabla_{\boldsymbol{Q}})^{2}
                        +V_{\text{cage}}(\boldsymbol{Q}),
\end{align}
where $V_{\text{cage}}(\boldsymbol{Q})$ is the cage potential and $M$ the mass of the ion. Hybridization modified by the ion displacement 
$\boldsymbol{Q}$ is expressed by the overlap integral between a plane wave $(1/\sqrt{\Omega})\text{e}^{\text{i}\boldsymbol{k}\cdot\boldsymbol{r}}$ 
and the localized orbitals of the magnetic ion, where $\Omega$ is the volume of the system. Without the spin orbit interaction, 
$\ell$-wave electron orbitals, $\psi_{\ell m}(\boldsymbol{r})\equiv u_{\ell}(r)Y_{\ell m}(\hat{\boldsymbol{r}})$, satisfy 
the following $\mathrm{Schr\ddot{o}dinger}$ equation,  
\begin{align}
\biggl(-\frac{\hbar^{2}\nabla^{2}}{2m}+V_{\text{nucleus}}(r) - \varepsilon_{\ell}\biggr)\psi_{\ell m}(\boldsymbol{r})=0,\label{sh_eq}
\end{align}
where the potential $V_{\text{nucleus}}(r)$ is assumed to be isotropic and $m$ is the mass of electron. When the ion is 
shifted from the origin by $\boldsymbol{Q}$, the relative coordinate from the ion to electron coordinate $\boldsymbol{r}$ 
is given by $\boldsymbol{r}-\boldsymbol{Q}$. Therefore the overlap integrals are expressed by
\begin{align}
V_{(\boldsymbol{k};\ell m)}(\boldsymbol{Q})=\frac{g}{\sqrt{\Omega}}\int d^{3}\boldsymbol{r}e^{-\text{i}\boldsymbol{k}\cdot\boldsymbol{r}}
\psi_{\ell m}(\boldsymbol{r}-\boldsymbol{Q}),
\end{align}
where $g$ is a coupling constant. With these integrals, hybridization terms modified by the ion displacement are given as
\begin{align}
\mathit{H}_{\text{hyb}}=\sum_{\boldsymbol{k}\sigma}\sum_{m}
\bigl[V_{(\boldsymbol{k};\ell m)}(\boldsymbol{Q})c^{\dagger}_{\boldsymbol{k}\sigma}f_{\ell m\sigma}+\text{h.c.}\bigr],
\end{align}
where $f_{\ell m\sigma}$ $(f^{\dagger}_{\ell m\sigma})$ are the annihilation (creation) operators for the localized $\ell$-wave orbitals. 
We expand the overlap integrals with respect to the ion displacement $\boldsymbol{Q}$. In this paper we focus on the first order terms of 
$\boldsymbol{Q}$. The hybridization terms expanded within $O(\boldsymbol{Q})$ are given as
\begin{align}
  \mathit{H}_{\text{hyb}}
=&\sum_{km\sigma}\biggl\{
   V_{\ell  }c^{\dagger}_{\ell   m  \sigma}(k)f_{\ell m\sigma}+\sum^{1}_{m'=-1}q^{*}_{1,m'}\notag\\
\times&\ \biggl[V^{(m,m')}_{\ell+1}c^{\dagger}_{\ell+1,m+m'\sigma}(k)
              f_{\ell m\sigma}\notag\\
 &+V^{(m,m')}_{\ell-1}c^{\dagger}_{\ell-1,m+m'\sigma}(k)
              f_{\ell m\sigma}\biggr] +\text{h.c.} \biggr\},\label{hybridization}
\end{align}
where $V_{\ell}$, $V^{(m,m')}_{\ell\pm1}$ and $q_{1,m}$ are defined in Appendix. It is obvious that the rotational symmetry 
$SO(3)$ is preserved in the generalized hybridization term (\ref{hybridization}). 

Under the simplest assumption that the ion has an $s$-wave electron orbital, the hybridization terms (\ref{hybridization})
are rewritten based on the Cartesian coordinate system as
\begin{align}
  \mathit{H}_{\text{hyb}} 
=&\ \ \ \sum_{k\sigma}\bigl\{V_{0}c^{\dagger}_{0\sigma}(k)f_{\sigma}+\text{h.c.}\bigr\}\notag\\
 &+\sum_{k\sigma}\sum_{i=x,y,z}\bigl\{\overline{V_{1}}c^{\dagger}_{1,i\sigma}(k)
              f_{\sigma}q_{i}+\text{h.c.}\bigr\},\label{hybridization2}
\end{align}
where $c_{0\sigma}$ $(c_{1,i\sigma})$ are the annihilation operators for the $s$ $(p_{i})$-wave components of the conduction electrons. 
It is verified easily that $V^{(0,0)}_{1}=V^{(0,\pm 1)}_{1}\equiv\overline{V_{1}}$ because of the rotational invariance.

For simplicity, we study the situation that the ion vibrates along one direction, say $x$-direction.
Using a one-dimensional harmonic potential $(1/2)M\omega^{2}Q^{2}$, we get the Hamiltonian for this study,
\begin{align}
    \mathit{H}
=&  \mathit{H}_{\text{c}}+\mathit{H}_{\text{hyb}}+\mathit{H}_{\text{local}},\label{full_Ham}\\
    \mathit{H}_{\text{c}}
=&  \sum_{k\sigma} \varepsilon(k)\bigl\{c^{\dagger}_{0\sigma}(k)c_{0\sigma}(k)
                                       +c^{\dagger}_{1\sigma}(k)c_{1\sigma}(k)\bigr\},\\
    \mathit{H}_{\text{hyb}} 
=&  \sum_{k\sigma}\bigl\{V_{0}c^{\dagger}_{0\sigma}(k)f_{\sigma}+\text{h.c.}\notag\\
 &                      +V_{1}c^{\dagger}_{1\sigma}(k)f_{\sigma}(a+a^{\dagger})+\text{h.c.}\bigr\},\\
    \mathit{H}_{\text{local}}
=&  \hbar\omega a^{\dagger}a
  + \varepsilon_{f}\sum_{\sigma}f^{\dagger}_{\sigma}f_{\sigma} + 
    Uf^{\dagger}_{\uparrow}f_{\uparrow}f^{\dagger}_{\downarrow}f_{\downarrow}\label{end},
\end{align}
where $\overline{V_{1}}/\sqrt{2}$ is set to $V_{1}$ and $a$ $(a^{\dagger})$ is the annihilation (creation) operator 
of the phonon. For the localized electron orbital, the annihilation (creation) operator is expressed by 
$f_{\sigma}$ $(f^{\dagger}_{\sigma})$ and $\epsilon_{f}$ is its energy given by eq. (\ref{sh_eq}) and $U$ the Coulomb interaction. 
When we restrict the isotropic degrees of freedom of the ionic vibrations to one particular direction, the $SO(3)$ symmetry is 
lowered to the discrete inversion symmetry; $r\rightarrow-r$ and $Q\rightarrow-Q$.

\subsection{Comments on the symmetries}\label{symmetry}
There are three relevant quantum numbers for the present model (\ref{full_Ham})$-$(\ref{end}). It is obvious that the total 
electron number $N^{\text{tot}}$ and the $z$ component of total spin $S^{\text{tot}}_{z}$ are conserved. The last one is 
related to the inversion symmetry mentioned in the previous subsection. The operation of the inversion $\mathcal{P}$ is 
described in the second-quantized notations,
\begin{align}
\mathcal{P}f_{\sigma}\mathcal{P}^{-1} &= f_{\sigma},\\
\mathcal{P}c_{0\sigma}(k)\mathcal{P}^{-1} &=  c_{0\sigma}(k),\\
\mathcal{P}c_{1\sigma}(k)\mathcal{P}^{-1} &= -c_{1\sigma}(k),\\
\mathcal{P}q\mathcal{P}^{-1} &= -q.
\end{align}
We characterize the inversion symmetry by the total parity $P$ which is defined 
by the sum between the number of $p$-channel conduction electrons $N_{p}$ and the number of harmonic phonons 
$N_{\text{ph}}$, namely $P=N_{p}+N_{\text{ph}}\equiv0$ or $1$ $(\text{mod}\ 2)$\cite{Dagotto}. 
This is based on the fact that, within $O(\boldsymbol{Q})$, phonon-assisted hopping between the impurity orbital and 
the $p$-channel conduction electrons has to accompany absorption or emission of one phonon. This quantum number 
plays a crucial role in the analyses of the energy spectra.

In the original model used by Yu and Anderson {(YA model)\cite{Yu_Anderson}}, it is assumed that a non-magnetic 
ion, which is trapped in a one-dimensional harmonic potential, vibrates in spinless conduction electrons. The first 
order process of the displacement induces scattering processes between the $s$-wave and $p$-wave conduction 
electrons. No localized electron orbital is introduced in the model. The original YA model is written as
\begin{align}
  \mathit{H}_{\text{YA}} 
=&\ \ \ \sum_{k   }\bigl\{\varepsilon_{0}(k)c^{\dagger}_{0}(k)c_{0}(k)+
                          \varepsilon_{1}(k)c^{\dagger}_{1}(k)c_{1}(k)\bigr\}\notag\\
 &+     \sum_{k,k'}\bigl\{V_{1}c^{\dagger}_{0}(k)c_{1}(k')(a+a^{\dagger})+\text{h.c.}
                   \bigr\}\notag\\
 &+\hbar\omega a^{\dagger}a,\label{original_YA}
\end{align}
where $c_{0}(k)$ $(c_{1}(k))$ are the annihilation operators for the spinless $s$-wave $(p$-wave$)$ 
conduction electrons with the dispersion relation $\varepsilon_{0}(k)$ $(\varepsilon_{1}(k))$.

In general, $\varepsilon_{0}(k)$ and $\varepsilon_{1}(k)$ are different. However if we assume that $\varepsilon_{0}(k)$ 
and $\varepsilon_{1}(k)$ are the same, an artificial symmetry comes into existence. 
The symmetry, which we call channel symmetry, is the invariance under the following transformation $\mathcal{I}$,
\begin{align}
\mathcal{I}c_{0}(k)\mathcal{I}^{-1} &= c_{1}(k),\\
\mathcal{I}c_{1}(k)\mathcal{I}^{-1} &= c_{0}(k).
\end{align}
With the channel symmetry, the Hamiltonian (\ref{original_YA}) can be decomposed into two detached parts by 
introducing the bonding and antibonding orbitals,
\begin{align}
c_{\text{e}}(k) = \frac{c_{0}(k)+c_{1}(k)}{\sqrt{2}},\\
c_{\text{o}}(k) = \frac{c_{0}(k)-c_{1}(k)}{\sqrt{2}},
\end{align}
and there is no matrix element between the two potential minima in contrast to the implicit assumption made by Yu and Anderson. 
Therefore the breaking of the channel symmetry is needed for the realization of Yu and Anderson type Kondo 
(YAK) effect. Matsuura and Miyake showed that potential scattering processes of the second order of the ion displacement 
play such a role\cite{Matsuura_Miyake}.

These discussions are also valid for the spinful fermion case. As for the present Hamiltonian (\ref{full_Ham}), the 
introduction of the localized impurity orbital serves as a source of breaking of the channel symmetry. Therefore 
the Hamiltonian (\ref{full_Ham}) at $U=0$ is one realization of the YA model extended to the spinful case.

\section{Numerical renormalization group approach}\label{section_3}
\subsection{Mapping to Wilson chains}
It is well known that for the ordinary impurity Anderson model the numerical renormalization group (NRG)
method is a powerful one to calculate energy spectra and various physical quantities 
with high accuracy\cite{Wilson,krishna}. We apply the NRG method to the present model by discretizing the
two continuous conduction bands in the logarithmic energy scales characterized by $\varLambda$. 
Both the $s$-wave and $p$-wave conduction bands can be discretized in the same way 
because the dispersion relation $\varepsilon(\boldsymbol{k})$ is assumed to have no angular dependence 
on $\boldsymbol{k}$. The discretized Hamiltonian is given by the two one-dimensional half-chains corresponding 
to the $s$-wave and $p$-wave conduction bands, which may be called Wilson chains,
\begin{align}
 &\mathit{H_{N}}=\varLambda^{\frac{N-1}{2}}\notag\\
 \times&\Biggl\{\ \sum^{N-1}_{n=0,\sigma}\varLambda^{-\frac{n}{2}}\xi_{n}
   \bigl[s^{\dagger}_{n,\sigma}s_{n+1,\sigma}
         +p^{\dagger}_{n,\sigma}p_{n+1,\sigma}+\text{h.c.}\bigr]\notag\\
 &\ +\sum_{\sigma}\bigl[  \widetilde{V_{0}}s^{\dagger}_{0,\sigma}f_{\sigma}
                         +\widetilde{V_{1}}p^{\dagger}_{0,\sigma}f_{\sigma}(a+a^{\dagger})
                         +\text{h.c.}\bigr]\notag\\
 &\ +\frac{\widetilde{U}}{2}\biggl(\sum_{\sigma}f^{\dagger}_{\sigma}f_{\sigma}-1\biggr)^{2}
    +\biggl(\widetilde{\varepsilon_{f}}
           +\frac{\widetilde{U}}{2}\biggr)\sum_{\sigma}f^{\dagger}_{\sigma}f_{\sigma}\notag\\
 &\ +\hbar\widetilde{\omega}a^{\dagger}a\Biggr\}.\label{Wilson_chain}
\end{align}
Here, for a constant density of states $\rho= 1/2D$ with the band width of $2D$ 
the $n$-th hopping matrix element $\xi_{n}$ is given by
\begin{align}
\xi_{n}=\frac{1-\varLambda^{-n-1}}{\sqrt{1-\varLambda^{-2n-1}}\sqrt{1-\varLambda^{-2n-3}}}.
\end{align}
All the parameters with tilde are multiplied by the constant factor $2/\{D(1+\varLambda^{-1})\}$.
From now on we treat the symmetric case, $2\varepsilon_{f}+U=0$, for simplicity. The Hamiltonian (\ref{Wilson_chain}) 
is block-diagonalized and each block is
characterized by the set of quantum numbers, $N^{\text{tot}}$, 
$S^{\text{tot}}_{z}$ and the parity $P$ as discussed above. We set various parameters for NRG calculations as follows; 
$\varLambda=3.0$, cutoff number of phonon excitations 50, band width $D=1.0$ and $M=15000$ states are kept 
at each NRG step. 

\begin{figure}
   \begin{center}
         \resizebox{80mm}{!}{\includegraphics{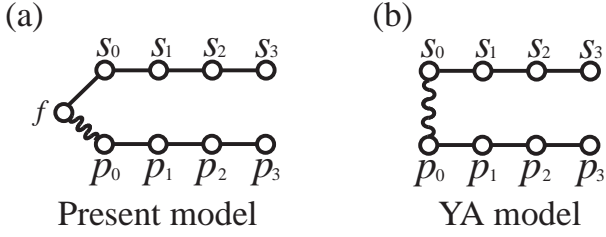}}
   \end{center}
   \caption{(Color online) (a) Graphical representation of the Hamiltonian (\ref{Wilson_chain}). The wavy line represents 
             phonon-assisted hopping while the straight lines usual hoppings. 
            (b) The YA model mapped to the Wilson chains.}\label{g_Anderson_and_YA_model}
\end{figure}

A graphical representation of the Hamiltonian (\ref{Wilson_chain}) is shown in Fig. \ref{g_Anderson_and_YA_model}(a). 
For comparison, the YA model mapped to the Wilson chains is also shown in Fig. \ref{g_Anderson_and_YA_model}(b). 
From these pictures, we understand that within the first order of ion displacement there is the channel symmetry 
in the original YA model unlike the present model when the same hopping matrix elements are used for both channels.

\subsection{Two-site problem\label{two-site}}
To investigate the effect of multi-phonon excitations, we consider the two-site Hamiltonian 
consisting of $f$ and $p_{0}$ sites in the Hamiltonian (\ref{Wilson_chain}),
\begin{align}
   \mathit{H}      _{\text{2-site}}
=& \mathit{H}^{(0)}_{\text{2-site}} + \mathit{H}^{(1)}_{\text{2-site}},\\
   \mathit{H}^{(0)}_{\text{2-site}}
=& V_{1}\sum_{\sigma}\bigl\{p^{\dagger}_{0,\sigma}f_{\sigma}+\text{h.c.}\bigr\}(a+a^{\dagger})\notag\\
 &+\hbar\omega a^{\dagger}a,\\
   \mathit{H}^{(1)}_{\text{2-site}}
=& \frac{U}{2}\biggl(\sum_{\sigma}f^{\dagger}_{\sigma}f_{\sigma}-1\biggr)^{2}.
\end{align}
For $U=0$, the two-site Hamiltonian is solved exactly 
by using the canonical transformation called Lang-Firsov transformation\cite{Lang-Firsov}, $e^{\text{i}S}$ with
\begin{align} 
S=\text{i}\lambda\sum_{\sigma}(-A^{\dagger}_{\sigma}A_{\sigma}
                        +B^{\dagger}_{\sigma}B_{\sigma})(a-a^{\dagger}),
\end{align}
where $A_{\sigma}$ and $B_{\sigma}$ are the annihilation operator for the antibonding and bonding orbitals, 
$A_{\sigma}=(1/\sqrt{2})(f_{\sigma} - p_{0,\sigma})$ and 
$B_{\sigma}=(1/\sqrt{2})(f_{\sigma} + p_{0,\sigma})$. The coupling constant is given by
$\lambda \equiv -V_{1}/\hbar\omega$. Then the transformed Hamiltonian is written as
\begin{align}
 &  \text{e}^{-\text{i}S}\mathit{H}^{(0)}_{\text{2-site}}\text{e}^{\text{i}S}\notag\\
=& -\frac{V^{2}_{1}}{\hbar\omega}
    \biggl\{\sum_{\sigma}(-A^{\dagger}_{\sigma}A_{\sigma}
                          +B^{\dagger}_{\sigma}B_{\sigma})\biggr\}^{2}+\hbar\omega a^{\dagger}a.
\end{align}
We obtain the doubly degenerate ground states $|\text{L}\bigl>$ and $|\text{R}\bigl>$, which may be called 
polaron doublet (PD), with the ground-state energy $E_{\text{PD}}$,
\begin{align}
|\text{L}\bigl> 
= &B^{\dagger}_{\uparrow}B^{\dagger}_{\downarrow}|\text{vac}\bigl>_{\text{el}}|\ell\bigl>_{\text{ph}}\notag\\
&\ \ \ \ \ \text{with}\ |\ell\bigl>_{\text{ph}} = \text{e}^{-2\lambda(a-a^{\dagger})}|0\bigl>_{\text{ph}}\label{L},\\
|\text{R}\bigl> 
= &A^{\dagger}_{\uparrow}A^{\dagger}_{\downarrow}|\text{vac}\bigl>_{\text{el}}|   r\bigl>_{\text{ph}}\notag\\
&\ \ \ \ \ \text{with}\ |   r\bigl>_{\text{ph}} = \text{e}^{ 2\lambda(a-a^{\dagger})}|0\bigl>_{\text{ph}}\label{R},\\
E_{\text{PD}}   = &-\frac{4V^{2}_{1}}{\hbar\omega}.
\end{align}
These two degenerate states are characterized by the expectation values of the ionic position $q$,
\begin{align}
&  \bigl<\text{L}|q    |\text{L}\bigl> 
= -\bigl<\text{R}|q    |\text{R}\bigl> = 2\sqrt{2}\lambda,\\
&  \bigl<\text{L}|q^{2}|\text{L}\bigl> 
=  \bigl<\text{R}|q^{2}|\text{R}\bigl> = \frac{16\lambda^{2}+1}{2}.
\end{align}
One can see that the PD is formed both for spinful and spinless cases and it plays a similar role as the localized 
spin for the usual spin Kondo effect.

In the following discussion, it is important to note that the electron parts of the doublet are written as
\begin{align}
 B^{\dagger}_{\uparrow}B^{\dagger}_{\downarrow}|\text{vac}\bigl>_{\text{el}}
=&|\text{LS}\bigl>_{\text{el}} + |\text{PS}\bigl>_{\text{el}},\\
 A^{\dagger}_{\uparrow}A^{\dagger}_{\downarrow}|\text{vac}\bigl>_{\text{el}}
=&|\text{LS}\bigl>_{\text{el}} - |\text{PS}\bigl>_{\text{el}},
\end{align}
where 
\begin{align}
       |\text{LS}\bigl>_{\text{el}}
&\equiv \frac{f^{\dagger}_{\uparrow}f^{\dagger}_{\downarrow}+p^{\dagger}_{0,\uparrow}p^{\dagger}_{0,\downarrow}}{2}
       |\text{vac}\bigl>_{\text{el}}
\end{align}
represents a local singlet with even parity while 
\begin{align}
       |\text{PS}\bigl>_{\text{el}}
&\equiv \frac{f^{\dagger}_{\uparrow}p^{\dagger}_{0,\downarrow}-f^{\dagger}_{\downarrow}p^{\dagger}_{0,\uparrow}}{2}
       |\text{vac}\bigl>_{\text{el}}
\end{align}
a pair singlet with odd parity. Similarly, the phonon parts of the doublet are written as
\begin{align}
  |\ell\bigl>_{\text{ph}}&=|\text{even}\bigl>_{\text{ph}} + |\text{odd}\bigl>_{\text{ph}},\\
  |   r\bigl>_{\text{ph}}&=|\text{even}\bigl>_{\text{ph}} - |\text{odd}\bigl>_{\text{ph}},
\end{align}
where $|\text{even}\bigl>_{\text{ph}}$ $\bigl(|\text{odd}\bigl>_{\text{ph}}\bigr)$ is a state which is a linear 
combination of states with even (odd) number of phonons. The parity of $|\text{even}\bigl>_{\text{ph}}$ 
$\bigl(|\text{odd}\bigl>_{\text{ph}}\bigr)$ is even (odd). Clearly $|\text{L}\bigl>$ and $|\text{R}\bigl>$ are not
eigenstates of the total parity, $P$.

The degeneracy of the PD is lifted by Coulomb interaction $\mathit{H}^{(1)}_{\text{2-site}}$, which may be treated 
by the perturbation theory. In fact energy shift from $E_{\text{PD}}$ is obtained by the following eigenvalue equation, 
\begin{align}
& \text{det}
\begin{pmatrix}
  \Delta E_{\text{PD}} - \bigl<\text{L}|\mathit{H}^{(1)}_{\text{2-site}}|\text{L}\bigl> 
&                       -\bigl<\text{L}|\mathit{H}^{(1)}_{\text{2-site}}|\text{R}\bigl>\\
                        -\bigl<\text{R}|\mathit{H}^{(1)}_{\text{2-site}}|\text{L}\bigl>  
& \Delta E_{\text{PD}} - \bigl<\text{R}|\mathit{H}^{(1)}_{\text{2-site}}|\text{R}\bigl>
\end{pmatrix}\notag\\
 =&\text{det}
\begin{pmatrix}
  \Delta E_{\text{PD}} - \frac{U}{4}                   
&                      - \frac{U}{4}\text{e}^{-8\lambda^{2}}\\
                       - \frac{U}{4}\text{e}^{-8\lambda^{2}}  
& \Delta E_{\text{PD}} - \frac{U}{4}
\end{pmatrix}=0.
\end{align}
Then within the first order perturbation of $U$, the eigenstates and eigenvalues are written as
\begin{align}
|\pm\bigl>          &= \frac{|\text{L}\bigl> \pm |\text{R}\bigl>}{\sqrt{2}},\\
E^{\pm}_{\text{PD}} &= E_{\text{PD}} + \frac{U}{4}\biggl(1 \pm \text{e}^{-8\lambda^{2}}\biggr).
\end{align}
It shows that Coulomb interaction works on the PD like a transversal magnetic field 
for a localized spin. The energy gap between $|+\bigl>$ and $|-\bigl>$, 
$\Delta_{\text{PD}}\equiv(U/2)\text{e}^{-8\lambda^{2}}$, is small when the electron-phonon couping $\lambda$ is strong, 
of the order of unity. Unlike $|\text{L}\bigl>$ and $|\text{R}\bigl>$, $|+\bigl>$ and $|-\bigl>$ are the eigenstates 
of the total parity. The parity of $|+\bigl>$ is even $(P=0)$ and that of $|-\bigl>$ is odd $(P=1)$.

\section{Results of the NRG calculations}\label{section_4}

\begin{figure}
   \begin{center}
         \resizebox{60mm}{!}{\includegraphics{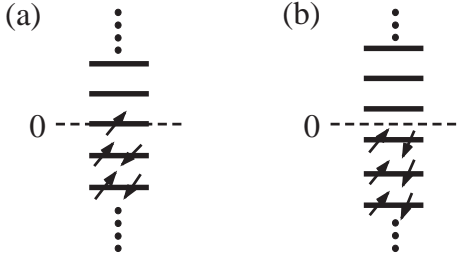}}
   \end{center}
   \caption{(Color online) Schematic one-particle energy levels of a single Wilson chain with (a) an odd number 
             of sites and (b) an even number of sites.}\label{picture_energy_level}
\end{figure}

\subsection{Classification of fixed points from energy spectra}\label{energy_level_analysis}
To begin with, we summarize the NRG calculations for the usual impurity Anderson model\cite{krishna}. After the mapping 
to a Wilson chain, the discretized Hamiltonian is composed of the impurity site connected to the single semi-infinite 
Wilson chain, which we will write as $f$ site and $s_{i}$ $(i=0\sim n)$ sites. The number of the NRG steps $n$ corresponds 
to the energy scale of the system. When $n$ is sufficiently large, energy flow approaches to the unique low-temperature 
fixed point.

The stable fixed point corresponds to the unitary limit of Kondo effect. In the limit, the low-energy spectrum is isomorphic 
to that of the Hamiltonian consisting of two parts, the singlet bond between $f$ and $s_{0}$ sites and the remaining free Wilson 
chain from $s_{1}$ to $s_{n}$. 

Concerning analysis of the fixed point, it is essential to note that the low-energy spectrum depends on even or odd of $n$. 
Eigenstates of the Wilson chain are specified by the set of quantum numbers, $[N^{\text{tot}}$, $S^{\text{tot}}_{z}]$. 
For a tight binding Hamiltonian with only off-diagonal matrix elements, single-particle energies are symmetric with respect to 
positive and negative energies. When $n$ is odd, the middle one of eigenenergies must be zero.  At half-filling Fermi energy 
should be also zero. Therefore, the ground states have four-fold degeneracy which are specified by the quantum numbers, $[n-1$, $0]$, 
$[n$, $\pm\frac{1}{2}]$ and $[n+1$, $0]$. For an even $n$, there is no eigenvalue at $E=0$ and therefore the ground state is 
unique with the quantum number $[n$, $0]$, see Fig. \ref{picture_energy_level}.
 
\begin{figure}
   \begin{center}
         \resizebox{80mm}{!}{\includegraphics{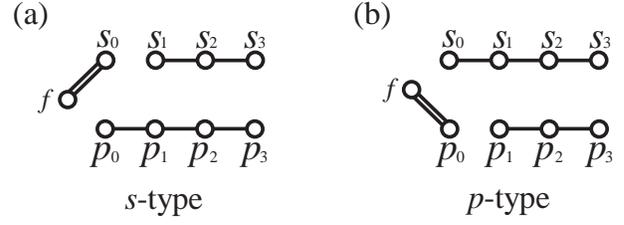}}
   \end{center}
   \caption{(Color online) Graphical representations of the two types of the energy spectra.}\label{site}
\end{figure} 

Now let us discuss the fixed points for the present model. After $n$ steps in the NRG, the system consists of $2n+3$ sites, 
the localized $f$ orbital and 
the two Wilson chains, $s_{0}$, $s_{1}$, $\cdots$, $s_{n}$ and $p_{0}$, $p_{1}$, $\cdots$, $p_{n}$.  Our numerical results 
by the NRG method show that there are only two stable fixed points: $s$- and $p$-type. The $s$-type fixed point schematically 
shown in Fig. \ref{site}(a) is characterized by the energy spectrum which is isomorphic to that of the system composed of 
tightly bound $f-s_{0}$ singlet and two independent Wilson chains whose length is different, $n$ sites for the $s$-chain 
and $n+1$ sites for the $p$-chain. It is important to include the parity as a quantum number, $[N^{\text{tot}}$, 
$S^{\text{tot}}_{z}$, $P]$. Concerning the $s$-type fixed point the phonon part does not play any important role and 
the total parity $P$ is simply determined by the number of electrons in the $p$-channel. From  Fig. \ref{site}(a) it 
is straightforward to see that the quantum numbers of the degenerate ground states are those given in 
Table \ref{ground_state_parity} depending on even or odd $n$. As an example, we show in Table \ref{energy_level_schK} 
a part of the low energy spectra of 
the $s$-type fixed point which correspond to the one particle excited states. 
The other fixed point defined as $p$-type is depicted in Fig. \ref{site}(b). 
The energy spectrum at low energies is isomorphic to that for the system with isolated $f-p_{0}$ singlet bond and two independent 
Wilson chains, $n+1$ sites for the $s$-chain and $n$ sites for the $p$-chain. One important non-trivial fact obtained by the present 
NRG calculations is that the quantum number for the part of $f-p_{0}$ singlet is always given 
by $[2$, $0$, $1]$, odd parity in the same way as the two-site problem discussed in \S\ref{two-site}. Therefore the quantum numbers 
of the ground states for the entire system are given by those shown in Table \ref{ground_state_parity}.

\begin{table}[ht!]
\begin{center}
\begingroup
\renewcommand{\arraystretch}{1.4}
\begin{tabular}{c|c|c|c}
\hline\hline
\multicolumn{2}{c|}{} & \multicolumn{2}{c}{NRG step $n$}\\
\cline{3-4}
\multicolumn{2}{c|}{} & even & odd \\
\hline
\multirow{6}{*}{fixed point} & \multirow{3}{*}{$s$-type} & $[2n+2$, $0$,              $0]$ & $[2n+2$, $0$,              $0]$ \\
                             &                           & $[2n+3$, $\pm\frac{1}{2}$, $1]$ & $[2n+3$, $\pm\frac{1}{2}$, $0]$ \\
                             &                           & $[2n+4$, $0$,              $0]$ & $[2n+4$, $0$,              $0]$ \\
\cline{2-4}
                             & \multirow{3}{*}{$p$-type} & $[2n+2$, $0$,              $1]$ & $[2n+2$, $0$,              $1]$ \\
                             &                           & $[2n+3$, $\pm\frac{1}{2}$, $1]$ & $[2n+3$, $\pm\frac{1}{2}$, $0]$ \\
                             &                           & $[2n+4$, $0$,              $1]$ & $[2n+4$, $0$,              $1]$ \\
\hline\hline
\end{tabular}
\endgroup
\end{center}
\caption{Sets of $[N^{\text{tot}},$ $S^{\text{tot}}_{z},$ $P]$ of the ground states at the $n$-th NRG step depending 
         on the types of the fixed points.}\label{ground_state_parity}
\end{table}

\begin{table}
\begin{center}
\begingroup
\renewcommand{\arraystretch}{1.4}
\begin{tabular}{c|c|c|c|c}
\multicolumn{5}{c}{\text{$s$-type fixed point}}\\
\hline\hline
 & \multicolumn{2}{c|}{$[1,$ $0,$ $0]$} & \multicolumn{2}{c}{$[1,$ $0,$ $1]$} \\
\hline
$\text{0th}$ & $0$        & $1$ & $0.800048$ & 2\\
$\text{1st}$ & $1.600095$ & $3$ & $2.400143$ & 2\\
$\text{2nd}$ & $1.695754$ & $2$ & $2.495802$ & 6\\
$\text{3rd}$ & $3.200191$ & $1$ & $2.997488$ & 2\\
\hline\hline
\end{tabular}
\endgroup
\end{center}
\caption{The low energy spectra of the $s$-type fixed point at an even number of the NRG steps
for the sets of the quantum numbers, $[1,$ $0,$ $0]$ and $[1,$ $0,$ $1]$. In this table, $N^{\text{tot}}$ 
is counted from the total number of sites. The energy eigenvalues and 
their degeneracy factors are tabulated up to the third excited states. The low energy spectra of the $p$-type fixed point 
can be obtained by the inversion of the total parity $P$.}\label{energy_level_schK}
\end{table}

Lastly, we discuss the unstable fixed points which separate the regions of the $s$- and $p$-type fixed points. Here, we briefly 
review the $2$-channel Kondo model. The model consists of a local moment coupled antiferromagnetically with two types of 
conduction electrons\cite{Cragg,Pang,Affleck}. The NRG calculations show that 2-channel Kondo effect is realized when the two 
antiferromagnetic coupling constants, $J_{a}$ and $J_{b}$, are perfectly equal\cite{Pang}. The energy spectra obtained from 
the NRG calculations are summarized in Refs. \ref{Affleck} and \ref{Cox_Zaw}. 
The low-energy spectrum of the $2$-channel Kondo model with 
$J_{a}=J_{b}$ shows $SO(5)$ rotational symmetry, in the sense that the degeneracy factors of the eigenstates are subject to 
dimensions of its irreducible representations.

Returning back to the present Hamiltonian, it is confirmed that the spectra of low-energy eigenvalues on the unstable 
fixed points are identical to those obtained for the $2$-channel Kondo model with $J_{a}=J_{b}$, see 
Table \ref{2chK_correspondence_relation}. Therefore, we describe these unstable fixed points as the $2$-channel Kondo 
($2$-chK) ones.

\begin{table}
\begin{center}
\begingroup
\renewcommand{\arraystretch}{1.4}
\begin{tabular}{c|ccc}
\hline\hline
$\text{Present model}$ & \multicolumn{3}{c}{\text{$2$-channel Kondo model}} \\
\hline
$[Q,$ $S^{\text{tot}}_{z},$ $P]$ & $(Q,$ $S,$ $S_{\text{ch}})$ & $SO(5)$ & $E_{\text{NRG}}$ \\
\hline
$[0,$ $\frac{1}{2},$ $1]$ & $(0,$ $\frac{1}{2},$ $0)$ &  1 & $0$ \\
\hline
$[\pm1,$ $0,$ $0]$ & \multirow{2}{*}{$(\pm1,$ $0,$ $\frac{1}{2})$} & \multirow{2}{*}{4} & \multirow{2}{*}{$0.125$} \\
$[\pm1,$ $0,$ $1]$ &                                       &                                               \\
\hline
$[\pm2,$ $\frac{1}{2},$ $0]$  & $(\pm2,$ $\frac{1}{2},$ $0)$ & \multirow{3}{*}{5} & \multirow{3}{*}{0.505}   \\
$[0,$ $\frac{1}{2},$ $0]^{*}$ & \multirow{2}{*}{$(0,$ $\frac{1}{2},$ $1)$}        &                        & \\
$[0,$ $\frac{1}{2},$ $1]$     &                                           &                        & \\
\hline
$[\pm1,$ $0,$ $0]$  & \multirow{2}{*}{$(\pm1,$ $1,$ $\frac{1}{2})$} & \multirow{2}{*}{$4$} & \multirow{2}{*}{$0.637$} \\
$[\pm1,$ $0,$ $1]$  &                                       &                      &                          \\
\hline
$[0,$ $\frac{1}{2},$ $1]$ & $(0,$ $\frac{3}{2},$ $0)$ & $1$ & $1.013$ \\
\hline\hline
\end{tabular}
\endgroup
\end{center}
\caption{
Correspondence relation of the low energy eigenstates of the $2$-chK fixed point between the present model and 
the $2$-channel Kondo model at an even number of the NRG steps. In this table, $Q$ is defined by the difference between 
$N^{\text{tot}}$ and the total number of sites. Spin degeneracy is not counted in this table. The right part of the table 
is from Table 12 in Ref. \ref{Cox_Zaw} with spin $S$ and channel spin $S_{\text{ch}}$ for the 2-channel Kondo model. *There 
are doubly degenerate states.}\label{2chK_correspondence_relation}
\end{table}

\subsection{Phase diagram of the generalized Anderson model for a vibrating magnetic ion}

\begin{figure}
   \begin{center}
      \resizebox{80mm}{!}{\includegraphics{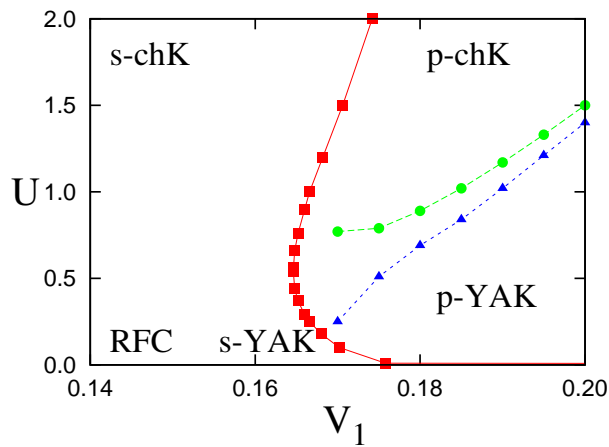}}
   \end{center}
   \caption{(Color online) The phase diagram in the parameter space of $V_{1}$ and $U$ for $V_{0}=\omega=0.2$. On the boundary, 
             the energy spectrum is of the $2$-chK type. The left region is characterized by the $s$-type energy spectra and 
             the right one by the $p$-type one. In the region between the green and blue dashed lines, the plateau of 
             $S_{\text{imp}}$ at $k_{\text{B}}\log2$ disappears.}\label{phase_diagram}
\end{figure}

From the energy spectra obtained by the NRG calculations, we determine the ground-state phase diagram which is shown 
in Fig. \ref{phase_diagram}. The parameter space is divided into two distinct areas. The left part is characterized 
by the $s$-type energy spectrum and the right part by the $p$-type one. On the boundary, the energy spectrum is of 
the $2$-chK type. On the line of $U=0$, the $s$-type energy spectrum always appears. Therefore the line of $2$-chK 
fixed points does not merge together to the $U=0$ line but approaches it asymptotically.

The NRG calculations reveal that physical properties of the system at finite temperatures are sometimes quite
different even if two different sets of parameters belong to the region of the same fixed point at zero temperature. 
Therefore, in addition to the analysis of the low temperature energy spectra, we have to 
investigate finite temperature behaviors of physical quantities at each point of the parameter space.

First, we focus on the left side of the $2$-chK line. On the line of $U=0$ this model is equivalent to the spinful YA model. 
In the weak electron-phonon coupling regime main source of kinetic energy comes from the hybridization with 
the $s$-wave conduction electrons and the phonon-assisted hybridization with the $p$-wave channel is a weak perturbation.
In this paper we characterize this regime by the renormalized Fermi chain (RFC). When $V_{1}$ increases, the PD is formed and 
the degeneracy is lifted at low temperatures by the Yu and Anderson type Kondo effect, expressed as $s$-YAK 
in Fig. \ref{phase_diagram}. Now, we gradually increase Coulomb interaction $U$, starting from the parameter region 
where temperature dependence of physical quantities is of the RFC type. The impurity begins to behave as a localized spin 
and  then the spin is screened mainly by the $s$-wave conduction electrons. This regime is called $s$-channel Kondo ($s$-chK) 
in Fig. \ref{phase_diagram}. In the parameter region between RFC and $s$-chK, the essence of the physics can be understood 
based on the local Fermi liquid theory\cite{Nozieres,Yamada_Yoshida}.

Next, we discuss the right part of the phase diagram. Concerning the strong $U$ regime, it is easy to understand that the local moment 
is screened mainly by the $p$-wave conduction electrons, which is called $p$-channel Kondo ($p$-chK). When $U$ is not important 
compared with $V_{1}$, the picture of the PD is valid, while the temperature dependence of physical quantities is different from that 
in the $s$-YAK parameter region in addition to the difference in the energy spectra. We denote this region as $p$-YAK 
in Fig. \ref{phase_diagram}. In the following subsections, we will discuss details of physical properties for various parameters, 
including the difference between the $s$-YAK and $p$-YAK.

\subsection{Three-site problem\label{3_site_problem}}
\begin{figure}
   \begin{center}
         \resizebox{80mm}{!}{\includegraphics{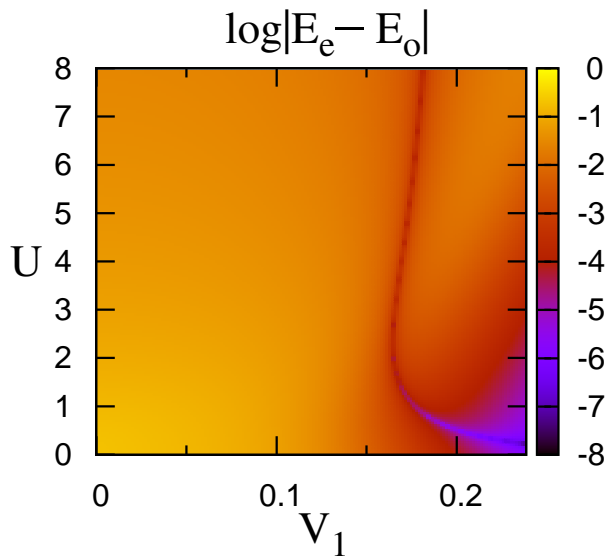}}
   \end{center}
   \caption{(Color online) The line of level crossing in the parameter space of $V_{1}$ and $U$ obtained from the three-site 
             two-electron problem for $V_{0}=\omega=0.2$. The color grade corresponds to the magnitude of 
             $\log|E_{\text{e}}-E_{\text{o}}|$, where $E_{\text{e}}$ and $E_{\text{o}}$ are the ground state energies 
             with even and odd parity, respectively. In the left part of the deep blue line, $E_{\text{e}}$ is lower than $E_{\text{o}}$. 
             }\label{3_site_pha_dia}
\end{figure}
\begin{figure}
   \begin{center}
         \resizebox{80mm}{!}{\includegraphics{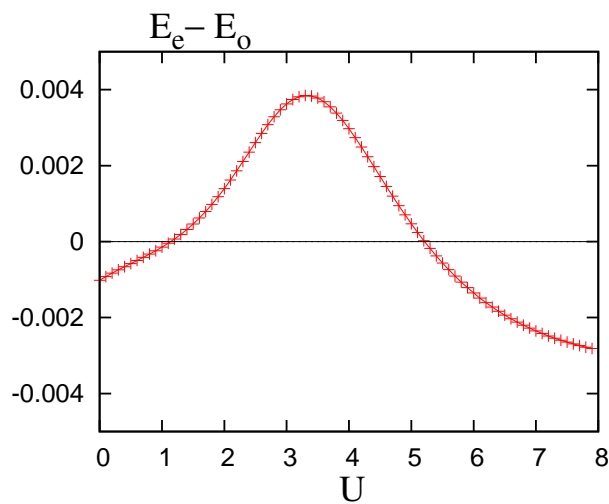}}
   \end{center}
   \caption{(Color online) $U$ dependence of the difference between $E_{\text{e}}$ and $E_{\text{o}}$ for 
             the three-site problem with two electrons: $V_{0}=\omega=0.2$ and $V_{1}=0.174$.}\label{energy_crossing}
\end{figure}
\begin{figure}
   \begin{center}
         \resizebox{80mm}{!}{\includegraphics{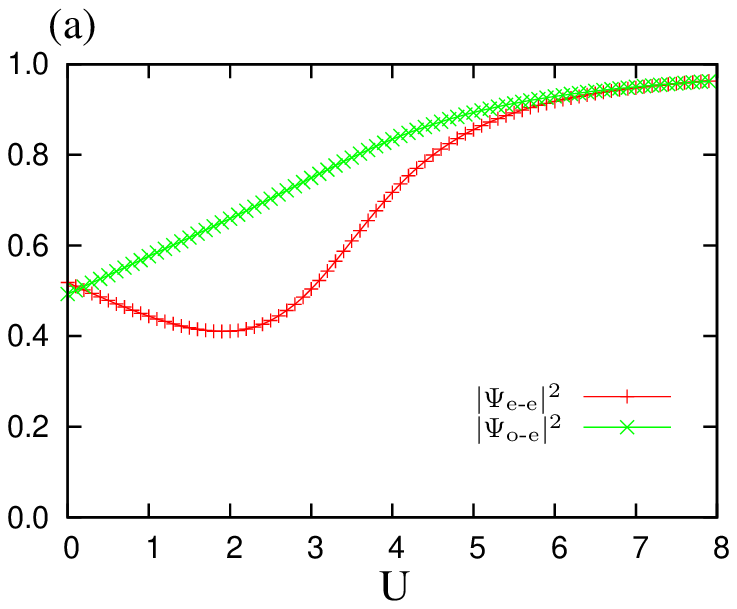}}\\
         \resizebox{80mm}{!}{\includegraphics{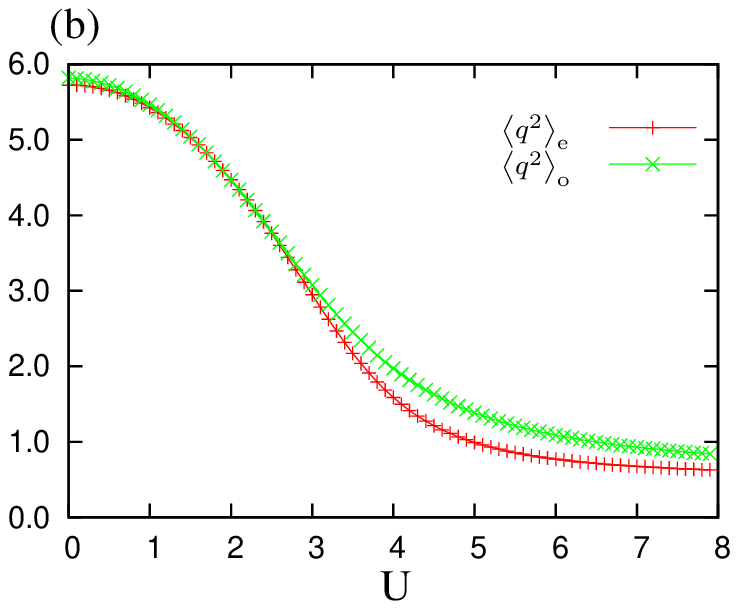}}
   \end{center}
   \caption{(Color online) (a) $U$ dependence of $|\Psi_{\text{e-e}}|^{2}$ and $|\Psi_{\text{o-e}}|^{2}$ obtained for 
             the three-site two-electron problem with $V_{0}=\omega=0.2$ and $V_{1}=0.174$. 
             (b) $U$ dependence of $\bigl<q^{2}\bigr>_{\text{e}}$ and $\bigl<q^{2}\bigr>_{\text{o}}$, which are 
             the expectation values of the square of the ion displacement calculated from $\Psi_{\text{even}}$ and 
             $\Psi_{\text{odd}}$.}\label{ProbAmp}
\end{figure}
At this point, it is instructive to consider the three-site problem, which consists of $f$, $s_{0}$ and $p_{0}$ sites. 
The Hamiltonian is given by
\begin{align}
   \mathit{H}_{\text{3-site}}
=&\ \ \ \ V_{0}\sum_{\sigma}\bigl\{s^{\dagger}_{\sigma}f_{\sigma}+\text{h.c.}\bigr\}\notag\\
 &+V_{1}\sum_{\sigma}\bigl\{p^{\dagger}_{\sigma}f_{\sigma}+\text{h.c.}\bigr\}(a+a^{\dagger})
  +\hbar\omega a^{\dagger}a\notag\\
 &+ \frac{U}{2}\biggl(\sum_{\sigma}f^{\dagger}_{\sigma}f_{\sigma}-1\biggr)^{2},
\end{align}
where the site suffix $"0"$ is suppressed. It is easy to obtain the exact eigenvalues and eigenstates  by using 
the exact diagonalization method. In the following examples the number of phonon excitations is kept up to 
$50$ and $V_{0}$ and $\omega$ are fixed to $0.2$.

Figure \ref{3_site_pha_dia} shows the magnitude of $\log|E_{\text{e}}-E_{\text{o}}|$ in the two-dimensional parameter space 
of $V_{1}$ and $U$, where $E_{\text{e}}$ $(E_{\text{o}})$ is the ground state energies with even and odd parity. 
In the figure there is a line of level crossing between $E_{\text{e}}$ and $E_{\text{o}}$. In the left (right) part 
of the line, $E_{\text{e}}$ $(E_{\text{o}})$ is lower than $E_{\text{o}}$ $(E_{\text{e}})$. 
Figure \ref{energy_crossing} shows $U$ dependence of the difference between $E_{\text{e}}$ and $E_{\text{o}}$ 
with $V_{1}$ fixed to $0.174$. Clearly there is similarity between Fig. \ref{phase_diagram} and Fig. \ref{3_site_pha_dia}.

From the view point of parity, every component of wave functions is classified into four groups, $\Psi_{\text{e-e}}$, 
$\Psi_{\text{e-o}}$, $\Psi_{\text{o-e}}$ and $\Psi_{\text{o-o}}$. $\Psi_{\text{e-o}}$, for example, means a linear combination
of the components whose electron parts have even parity and phonon parts odd parity. Clearly an eigenstate with even total parity
is written as
\begin{align}
\Psi_{\text{even}} &= \Psi_{\text{e-e}}+\Psi_{\text{o-o}},
\end{align}
while one with odd total parity as
\begin{align}
\Psi_{\text{odd}}  &= \Psi_{\text{o-e}}+\Psi_{\text{e-o}}.
\end{align}
Concerning the three-site two-electron problem, the bases for a given parity are listed in Table \ref{three_site_base}.

Figure \ref{ProbAmp}(a) shows $U$ dependence of the amplitudes, $|\Psi_{\text{e-e}}|^{2}$ and $|\Psi_{\text{o-e}}|^{2}$, 
for $V_{1}=0.174$. Because the total wave functions are normalized, $|\Psi_{\text{o-o}}|^{2}=1-|\Psi_{\text{e-e}}|^{2}$ 
and $|\Psi_{\text{e-o}}|^{2}=1-|\Psi_{\text{o-e}}|^{2}$. In Fig. \ref{ProbAmp}(b) we show the expectation values of the square of the ion 
displacement, $\bigl<q^{2}\bigr>_{\text{e}}$ and $\bigl<q^{2}\bigr>_{\text{o}}$, which are calculated from $\Psi_{\text{even}}$ and 
$\Psi_{\text{odd}}$, respectively. 

We start from the two-site problem for the non-interacting case, namely $V_{0}=U=0$, where the ground states $|+\bigl>$ and $|-\bigl>$ 
or $|\text{L}\bigl>$ and $|\text{R}\bigl>$ are doubly degenerate. When $V_{0}$ is introduced, we expect that the kinetic energy favors 
the configurations which belong to $\Psi_{\text{e-e}}$. Actually, the amplitude $|\Psi_{\text{e-e}}|^{2}$ shown in Fig. \ref{ProbAmp}(a) 
is larger than $0.5$ at $U=0$. As we have discussed in \S\ref{two-site} the Coulomb interaction $U$ favors the state with odd total parity.
The competition leads to the level crossing at small $U$.

\begin{table}
\begin{center}
\begingroup
\renewcommand{\arraystretch}{1.4}
\begin{tabular}{c|c|c}
\hline\hline
& \multicolumn{2}{c}{Parity $P$}\\
\cline{2-3}
& 0 & 1 \\
\hline
\multirow{3}{*}{electron} 
& $s^{\dagger}_{\uparrow}s^{\dagger}_{\downarrow}|\text{vac}\bigr>_{\text{el}}$ & \\
& $f^{\dagger}_{\uparrow}s^{\dagger}_{\downarrow}|\text{vac}\bigr>_{\text{el}}$,
  $s^{\dagger}_{\uparrow}f^{\dagger}_{\downarrow}|\text{vac}\bigr>_{\text{el}}$ 
& $s^{\dagger}_{\uparrow}p^{\dagger}_{\downarrow}|\text{vac}\bigr>_{\text{el}}$, 
  $p^{\dagger}_{\uparrow}s^{\dagger}_{\downarrow}|\text{vac}\bigr>_{\text{el}}$ \\
& $f^{\dagger}_{\uparrow}f^{\dagger}_{\downarrow}|\text{vac}\bigr>_{\text{el}}$, 
  $p^{\dagger}_{\uparrow}p^{\dagger}_{\downarrow}|\text{vac}\bigr>_{\text{el}}$  
& $f^{\dagger}_{\uparrow}p^{\dagger}_{\downarrow}|\text{vac}\bigr>_{\text{el}}$, 
  $p^{\dagger}_{\uparrow}f^{\dagger}_{\downarrow}|\text{vac}\bigr>_{\text{el}}$ \\
\cline{1-3}
\multirow{2}{*}{phonon} 
& \multirow{2}{*}{$|n\bigr>_{\text{ph}}$ $(n=$ \text{even}$)$} & \multirow{2}{*}{$|n\bigr>_{\text{ph}}$ $(n=$ \text{odd}$)$}\\
& & \\ 
\hline\hline
\end{tabular}
\endgroup
\end{center}
\caption{Bases for electron and phonon parts of the three-site two-electron problem classified by the parity $P$.}\label{three_site_base}
\end{table}

In the narrow region of $V_{1}$, the ground state makes the second level crossing at larger $U$. In the strongly correlated regime, 
a local moment is formed in the $f$-orbital, which should be eventually screened by the conduction electrons. The expectation values 
of square of the ion displacement are suppressed and become close to the amplitude of zero point fluctuations for both $\Psi_{\text{even}}$ 
and $\Psi_{\text{odd}}$ as shown in Fig. \ref{ProbAmp}(b). In this situation, the phonon parts of the wave functions are dominated by 
components with even phonon numbers. This fact is clearly seen in Fig. \ref{ProbAmp}(a). Therefore the total parity is predominantly 
determined by the nature of singlet formation, even parity for the $f-s$ singlet while odd parity for the $f-p$ singlet.
The upper critical $U$ shown in Fig. \ref{energy_crossing} corresponds to the point where the two screening mechanisms compete 
in the same way as the $2$-chK fixed point shown in the phase diagram for the bulk system, Fig. \ref{phase_diagram}.

The line of level crossing shown in Fig. \ref{3_site_pha_dia} is qualitatively very similar to the phase boundary obtained by the NRG 
calculations, Fig. \ref{phase_diagram}. We would like to point out that the three-site Hamiltonian corresponds to the initial step of 
the NRG procedure and the ground state of the two-electron problem has the different parity depending on the nature of the singlet bond 
as discussed in \S\ref{energy_level_analysis}, see also Table \ref{ground_state_parity}.

\subsection{Strong correlation regime $(U\gg\omega\gg T)$}
\begin{figure}
   \begin{center}
         \resizebox{80mm}{!}{\includegraphics{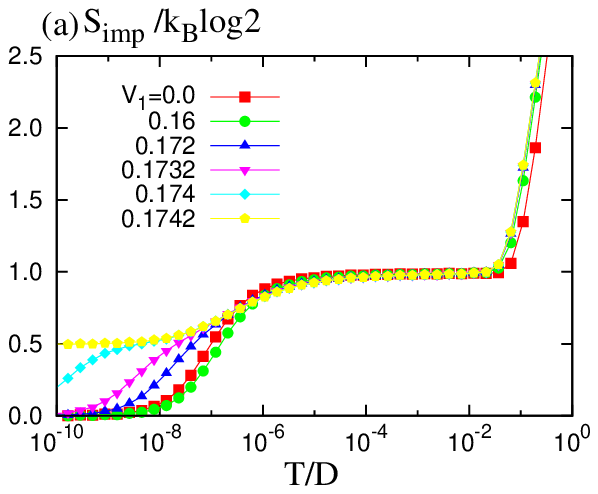}}\\
         \resizebox{80mm}{!}{\includegraphics{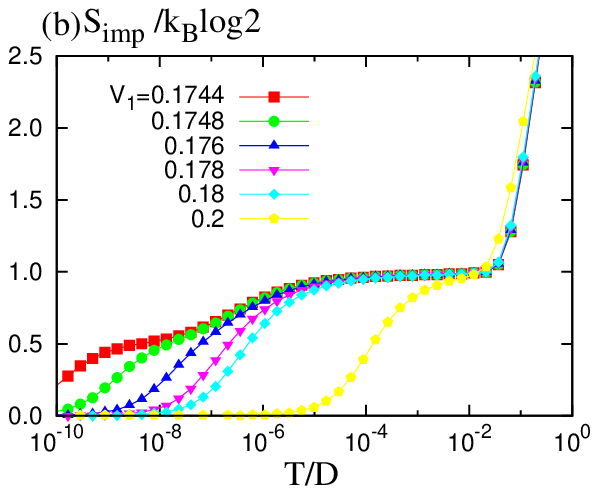}}
   \end{center}
   \caption{(Color online) Temperature dependence of entropy in the strong correlation regime. The used parameters are 
             $V_{0}=\omega=0.2$ with the symmetric condition $U=-2\varepsilon_{f}=2.0$. (a) For $V_{1}<0.1742$, the 
             $s$-chK effect is observed. (b) When $V_{1}$ is larger than the critical value the local moment is mainly 
             screened by the $p$-wave conduction electrons, corresponding to the $p$-chK effect. At $V_{1}=0.1742$, we observe 
             signature of the $2$-channel Kondo effect.}\label{Strong_Correlation_regime_entropy}
\end{figure}
\begin{figure}
   \begin{center}
      \begin{tabular}{cc}
         \resizebox{80mm}{!}{\includegraphics{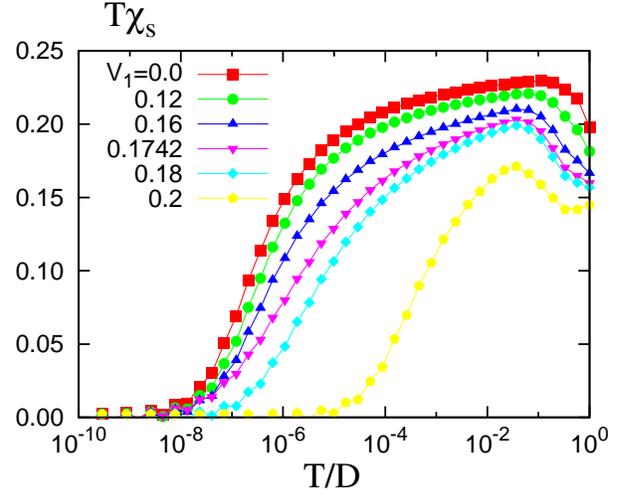}}
      \end{tabular}
   \end{center}
   \caption{(Color online) Temperature dependence of spin susceptibility $T\chi_{s}$ in the strong correlation regime, 
             $V_{0}=\omega=0.2$ with the symmetric condition $U=-2\varepsilon_{f}=2.0$. When $V_{1}$ is relatively 
             small, $T\chi_{s}$ increases close to 0.25 which corresponds to the local moment limit. 
             With increasing $V_{1}$, phonon-assisted hybridization suppresses the enhancement of $T\chi_{s}$.
           }\label{Strong_Correlation_regime_spin}
\end{figure}
Firstly, we discuss physical properties in the strong $U$ regime where the relation, $U\gg\omega\gg T$, holds. 
In this regime multi phonon excitation processes are not important. Because of the strong $U$, the localized orbital 
behaves as a local moment which is coupled with the $s$-wave and $p$-wave conduction electrons. 
With lowering temperature, the local moment is screened by the conduction electrons and the main screening channel 
changes depending on relative strength between $V_{0}$ and $V_{1}$. In the same way as the conventional impurity Anderson model 
is mapped to the spin $1/2$ Kondo model\cite{SH_trans}, we consider the projected Hilbert space where the vacuum and 
doubly occupied states of the impurity orbital are prohibited. We treat both of the conventional hybridization term 
$V_{0}$ and the phonon-assisted one $V_{1}$ in the Hamiltonian (\ref{full_Ham}) as perturbation terms. After some 
calculations, a generalized Kondo model is obtained as
\begin{align}
  \mathit{H}_{\text{Kondo}} 
=& \sum_{k\sigma}\sum_{\ell=0,1} \varepsilon(k)c^{\dagger}_{\ell\sigma}(k)c_{\ell\sigma}(k)+\hbar\omega a^{\dagger}a
  +\mathit{H}_{\text{s-d}},\notag\\
   \mathit{H}_{\text{s-d}}
=&-{\sum_{kk'}}\sum_{\alpha\beta}\biggl\{J_{0,0}
   c^{\dagger}_{0\alpha}(k)\bigl[\boldsymbol{\sigma}\bigr]_{\alpha\beta}c_{0\beta}(k')\cdot\mathbf{S}\notag\\
 &+J_{1,0}~q\bigl[c^{\dagger}_{1\alpha}(k)\bigl[\boldsymbol{\sigma}\bigr]_{\alpha\beta}c_{0\beta}(k')\cdot\mathbf{S}
  +\text{h.c.}\bigr]\notag\\
 &+J_{1,1}~q^{2}c^{\dagger}_{1\alpha}(k)\bigl[\boldsymbol{\sigma}\bigr]_{\alpha\beta}c_{1\beta}(k')\cdot\mathbf{S}
    \biggr\},\label{Kondo}
\end{align}
where $\mathbf{S}=f^{\dagger}_{\alpha}[\boldsymbol{\sigma}]_{\alpha\beta}f_{\beta}$ with 
$\boldsymbol{\sigma}$ being the vector of Pauli matrices and $q=(a+a^{\dagger})/\sqrt{2}$. The coupling constants $J_{0,0}$, 
$J_{1,0}$ and $J_{1,1}$ are written as
\begin{align}
J_{0,0} \equiv&  V^{2}_{0}  \biggl\{\frac{1}{\varepsilon_{f}}-\frac{1}{\varepsilon_{f}+U}\biggr\},\notag\\
J_{1,0} \equiv&  V_{0}      \bigl(\sqrt{2}V_{1}\bigr) 
                            \biggl\{\frac{1}{\varepsilon_{f}}-\frac{1}{\varepsilon_{f}+U}\biggr\},\notag\\
J_{1,1} \equiv& 2V^{2}_{1}  \biggl\{\frac{1}{\varepsilon_{f}}-\frac{1}{\varepsilon_{f}+U}\biggr\}.\notag
\end{align}
The term proportional to $J_{1,0}$ corresponds to scattering processes from the $s$-wave conduction electrons to 
the $p$-wave ones and vice versa. Similarly, the term proportional to $J_{1,1}$ describes scattering processes within the 
$p$-wave conduction electrons. In the strong correlation regime, effects of electron-phonon coupling are suppressed 
by Coulomb interaction. Therefore we may regard the effects of phonon emission or absorption processes as a 
renormalization of the coupling constant by taking average over the phonon degree of
freedom,
\begin{align}
   \mathit{H}_{\text{s-d}}
\simeq
 &-{\sum_{kk'}}\sum_{\alpha\beta}\bigl\{J_{0,0}
   c^{\dagger}_{0\alpha}(k)\bigl[\boldsymbol{\sigma}\bigr]_{\alpha\beta}c_{0\beta}(k')\cdot\mathbf{S}\notag\\
 &+J_{1,1}\bigl<q^{2}\bigr>
          c^{\dagger}_{1\alpha}(k)\bigl[\boldsymbol{\sigma}\bigr]_{\alpha\beta}c_{1\beta}(k')\cdot\mathbf{S}
    \bigr\},\label{Kondo2}
\end{align}
where used is the fact that $\bigl<q\bigr>$, the thermal averages of $q$, vanishes. Even if the electron-phonon coupling 
is weak, $\bigl<q^{2}\bigr>$ is not zero due to the zero point fluctuations.

In Figs. \ref{Strong_Correlation_regime_entropy} and \ref{Strong_Correlation_regime_spin} we use a set of parameters, 
$V_{0}=\omega=0.2$ with the symmetric condition $U=-2\varepsilon_{f}=2.0$. When $V_{1}$ is relatively small, 
the local moment is mainly screened by the $s$-channel conduction electrons, corresponding to 
the $s$-channel Kondo effect. Figure \ref{Strong_Correlation_regime_entropy}(a) shows that entropy of the impurity
site $S_{\text{imp}}$ is not affected so much by the $p$-wave conduction electrons for $V_{1}\leq0.16$. 
Spin susceptibility $T\chi_{s}$ shown in Fig. \ref{Strong_Correlation_regime_spin} reaches close to $0.25$ of 
the local moment limit because of strong $U$. At the critical $V_{1}(=0.1742)$, the energy spectra are characterized 
by the $2$-channel Kondo fixed point and accordingly the plateau of entropy at $(1/2)k_{\text{B}}\log2$ 
appears at low temperatures. When $V_{1}$ is larger than the critical value, a different type of Kondo effect takes 
place through screening by the $p$-wave conduction electrons as shown in Fig. \ref{Strong_Correlation_regime_entropy}(b). 
With further increase of $V_{1}$, effective Coulomb interaction at the impurity site is 
suppressed by the electron-phonon coupling. This effect of renormalization of effective Coulomb interaction appears 
as the higher Kondo temperatures in Fig. \ref{Strong_Correlation_regime_entropy}(b) and also as the suppression of $T\chi_{s}$ 
in Fig. \ref{Strong_Correlation_regime_spin}.

At the $2$-channel Kondo fixed point, the thermal average of square of ion displacement $\bigl<q^{2}\bigr>$ converges 
to $0.739$ at low temperatures. When we substitute the number into $\bigl<q^{2}\bigr>$, the ratio of 
$J_{1,1}\bigl<q^{2}\bigr>$ to $J_{0,0}$ is estimated to be
\begin{align}
\frac{J_{1,1}\bigl<q^{2}\bigr>}{J_{0,0}} \simeq 1.12.
\end{align}
The estimated value close to $1$ seems to be reasonable.

\subsection{Non-interacting case $(U=0)$}\label{non_int}
\begin{figure}
   \begin{center}
      \begin{tabular}{cc}
         \resizebox{80mm}{!}{\includegraphics{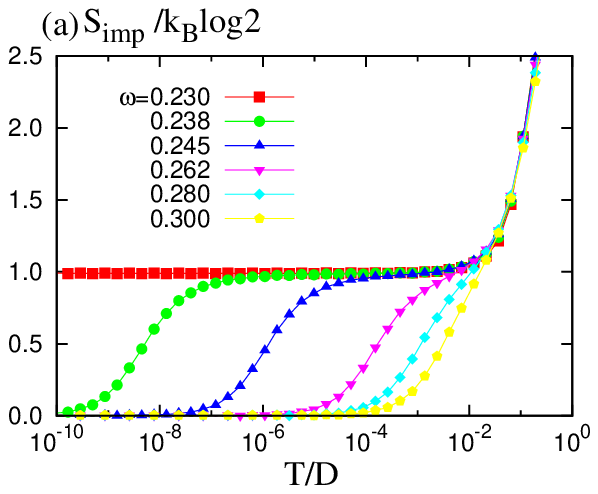}}\\
         \resizebox{82mm}{!}{\includegraphics{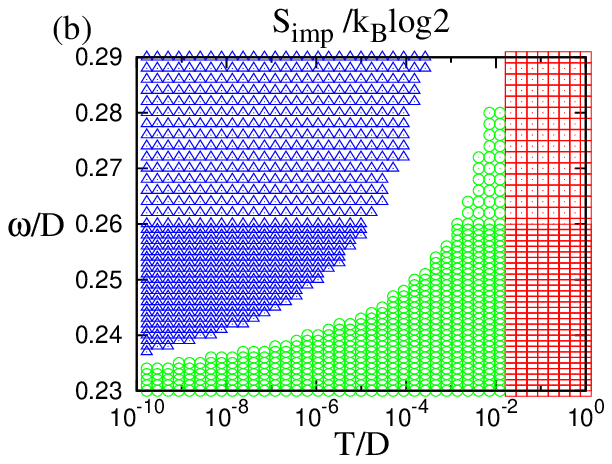}}
      \end{tabular}
   \end{center}
   \caption{(Color online) Temperature dependence of entropy for various phonon frequency $\omega$ with $V_{0}=V_{1}=0.2$ 
             and $U=0$. When $\omega$ is small, $S_{\text{imp}}$ shows a plateau at $k_{\text{B}}\log2$. 
             For $\omega<0.28$, the plateau due to the PD is suppressed in the same temperature dependence as the usual 
             Kondo effect. Results on the entropy for various $\omega$ are assembled in the parameter plane of $\omega$ 
             and $T$ in the lower panel. The red squares are the points where the entropy is larger than $1.1k_{\text{B}}\log2$ 
             and the green circles are for $0.9k_{\text{B}}\log2<S_{\text{imp}}<1.1k_{\text{B}}\log2$ and the blue 
             triangles for $S_{\text{imp}}<0.1k_{\text{B}}\log2$.}\label{entropy_Yu_and_Anderson}
\end{figure}

\begin{figure}
   \begin{center}
      \begin{tabular}{cc}
         \resizebox{80mm}{!}{\includegraphics{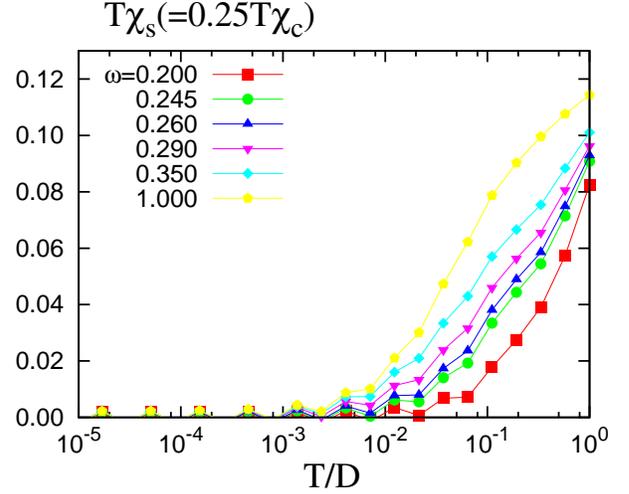}}
      \end{tabular}
   \end{center}
   \caption{(Color online) Temperature dependence of spin susceptibility $T\chi_{s}(=0.25T\chi_{c})$ for $V_{0}=V_{1}=0.2$ 
             and $U=0$. When $\omega$ is lowered, $T\chi_{s}(=0.25T\chi_{c})$ is suppressed since strong electron-phonon 
             coupling increases the binding energy of the PD.}\label{spin_Yu_and_Anderson}
\end{figure}

\begin{figure}
   \begin{center}
      \begin{tabular}{cc}
         \resizebox{80mm}{!}{\includegraphics{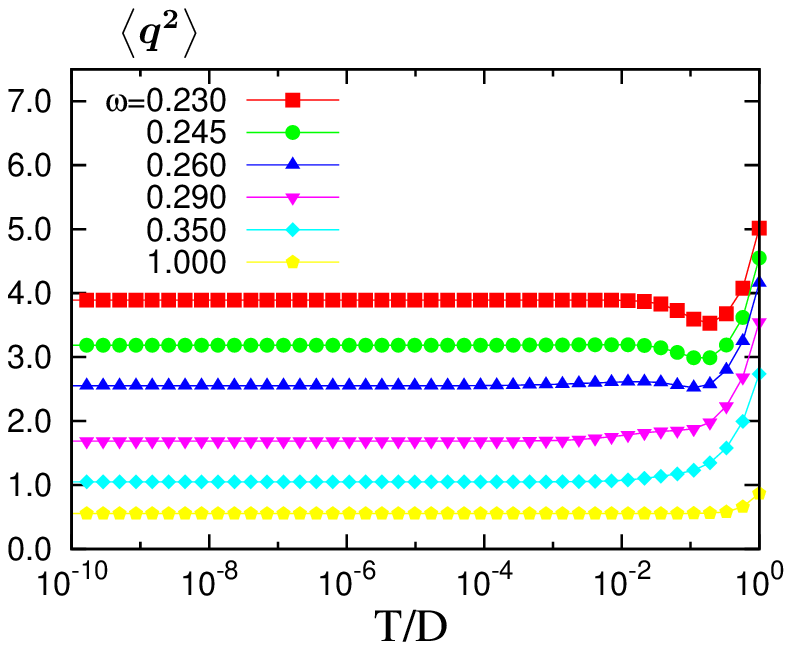}}
      \end{tabular}
   \end{center}
   \caption{(Color online) Thermal average of square of ion displacement, $\bigl<q^{2}\bigr>$, for $V_{0}=V_{1}=0.2$ and $U=0$.
             When $\omega$ becomes larger, $\bigl<q^{2}\bigr>$ asymptotically comes close to $0.5$ which corresponds to the zero point
             fluctuation.}\label{squared_q_Yu_and_Anderson}
\end{figure}
For discussions on the non-interacting case we fix the set of parameters $V_{0}=V_{1}=0.2$ and $U=0.0$. In this 
subsection we use frequency of phonon $\omega$ as a parameter to change the strength of electron-phonon 
coupling, which is an alternative to control $V_{1}$ with a fixed $\omega$.

Entropy of the impurity site $S_{\text{imp}}$ is shown in Fig. \ref{entropy_Yu_and_Anderson}(a) for various 
$\omega$. As $\omega$ decreases, the plateau of entropy at $k_{\text{B}}\log2$ develops owing to 
the formation of the PD. Figure \ref{entropy_Yu_and_Anderson}(b) summarizes results on the entropy in $\omega-T$ 
plane. Typical YAK behaviors are observed in the parameter range from $\omega=0.238$ to $0.280$. The middle of the blank 
area between the blue triangle and green circle regions corresponds to the Kondo temperature for the YAK effect.

In Fig. \ref{spin_Yu_and_Anderson} the spin (charge) susceptibility of the impurity, $T\chi_{s}$ $(T\chi_{c})$, is shown for 
various $\omega$. In general the relation $T\chi_{s}=0.25T\chi_{c}$ holds in the non-interacting cases. 
When temperature is lower than the binding energy of the PD, $E_{\text{PD}}$, spin and charge fluctuations at the impurity site 
are suppressed due to the formation of the PD which consists of the spin singlets of doubly occupied bonding or 
anti-bonding states. Therefore, unlike the usual spin Kondo effect temperature dependence of $T\chi_{s}$ ($T\chi_{c}$) is governed 
by $E_{\text{PD}}$ rather than the Kondo temperature of the YAK effect. 

Figure \ref{squared_q_Yu_and_Anderson} shows temperature dependence of the thermal average of square of ion displacement, 
$\bigl<q^{2}\bigr>$. For all $\omega$, $\bigl<q^{2}\bigr>$ converges to a constant at low temperatures and does not show 
any changes in the vicinity of the YAK temperature. Once the PD is formed the phonon states are almost frozen 
and given by the two distinct coherent states which are symmetrically shifted right and left from the center. It is interesting 
to note that $\bigl<q^{2}\bigr>$ for $\omega$=$0.230$, $0.245$ and $0.260$ show minima in the temperature range from $10^{0}$ 
to $10^{-2}$ which may correspond to $E_{\text{PD}}$. These behaviors may be interpreted as formation of the effective 
double-well potential for the ion displacement. When temperature is higher than $E_{\text{PD}}$, the dynamics of the ion is
governed by the original harmonic cage potential. On the other hand, below the energy scale of $E_{\text{PD}}$, the ion tunnels 
between the left and right minima corresponding to $|\text{L}\bigl>$ and $|\text{R}\bigl>$.

\subsection{Effect of finite $U$ in the weak electron-phonon coupling regime}
\begin{figure}
   \begin{center}
      \begin{tabular}{cc}
         \resizebox{80mm}{!}{\includegraphics{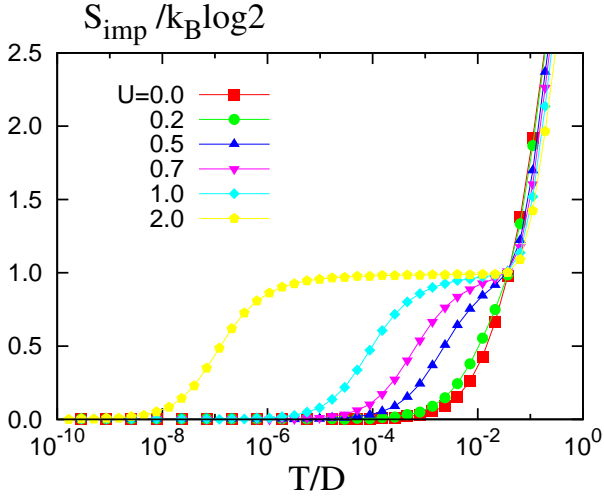}}
      \end{tabular}
   \end{center}
   \caption{(Color online) Temperature dependence of entropy in the weak electron-phonon coupling regime,  
             $V_{0}=2V_{1}=\omega=0.2$. When $U$ is increased, a plateau of entropy at 
             $k_{\text{B}}\log2$ appears because of formation of the local moment. At low temperatures 
             the local moment is screened mainly by the $s$-wave conduction electrons.}\label{entropy_weak_U}
\end{figure}

\begin{figure}
   \begin{center}
      \begin{tabular}{cc}
         \resizebox{80mm}{!}{\includegraphics{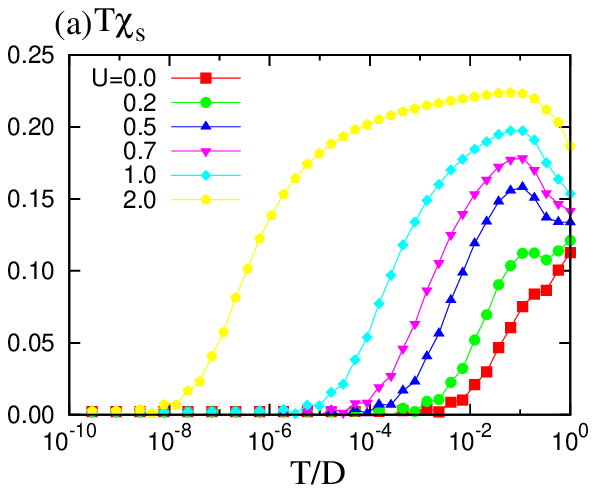}}\\
         \resizebox{80mm}{!}{\includegraphics{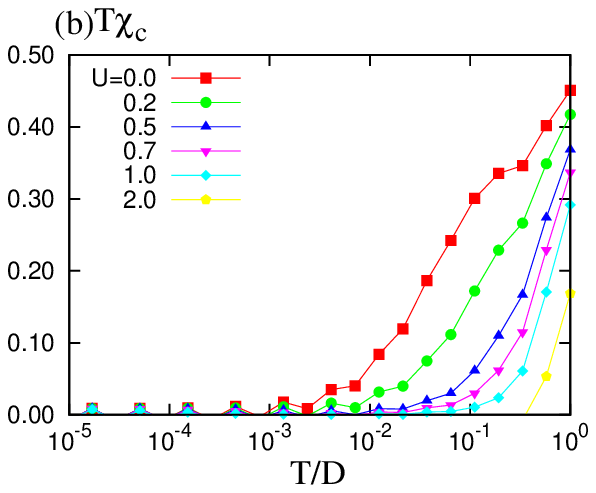}}
      \end{tabular}
   \end{center}
   \caption{(Color online) Temperature dependence of $T\chi_{s}$ shown in the upper panel 
             and $T\chi_{c}$ in the lower one in the weak electron-phonon coupling regime, $V_{0}=2V_{1}=\omega=0.2$. 
             The relation, $T\chi_{s}=0.25T\chi_{c}$, is valid for $U=0$. When $U$ is increased, $T\chi_{s}$ 
             is enhanced while $T\chi_{c}$ is suppressed.}\label{Strong_Correlation_regime_spin_charge}   
\end{figure}
Next we choose the set of parameters $V_{0}=2V_{1}=\omega=0.2$, where screening by the $s$-wave conduction 
electrons dominates over the $p$-wave ones. To investigate the effects of Coulomb interaction, we show temperature 
dependence of the physical quantities, the entropy of the impurity ion $S_{\text{imp}}$ in Fig. \ref{entropy_weak_U}, 
the spin susceptibility $T\chi_{\text{s}}$ in Fig. \ref{Strong_Correlation_regime_spin_charge}(a) and the charge susceptibility 
$T\chi_{\text{c}}$ in Fig. \ref{Strong_Correlation_regime_spin_charge}(b). When $U$ is small, $S_{\text{imp}}$ 
monotonically decays to zero. $T\chi_{s}$ and $T\chi_{c}$ also go to zero without any plateau. Therefore, we 
conclude that the physical properties correspond to the RFC regime in Fig. \ref{phase_diagram}. When $U$ increases, 
the impurity orbital begins to behave as a local moment. Actually, the plateau region in Fig. \ref{entropy_weak_U} 
is broadened with increasing $U$. Correspondingly, $T\chi_{s}$ is enhanced and shows a maximum close to $0.25$ while 
$T\chi_{c}$ is suppressed, which are signatures of a local moment. As already mentioned, the energy spectra are characterized 
by the $s$-type fixed point for any $U$. However the physical properties cross over from the RFC behaviors to the $s$-chK ones 
with increasing $U$.

\subsection{Effect of finite $U$ in the strong electron-phonon coupling regime\label{strong_e_p}}
\begin{figure}
   \begin{center}
      \begin{tabular}{cc}
         \resizebox{80mm}{!}{\includegraphics{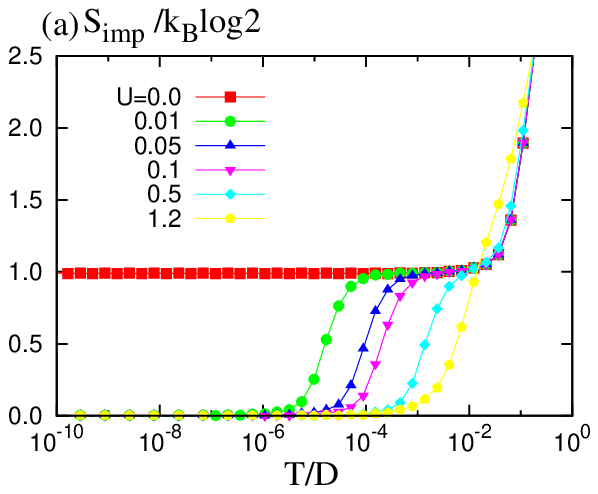}}\\
         \resizebox{80mm}{!}{\includegraphics{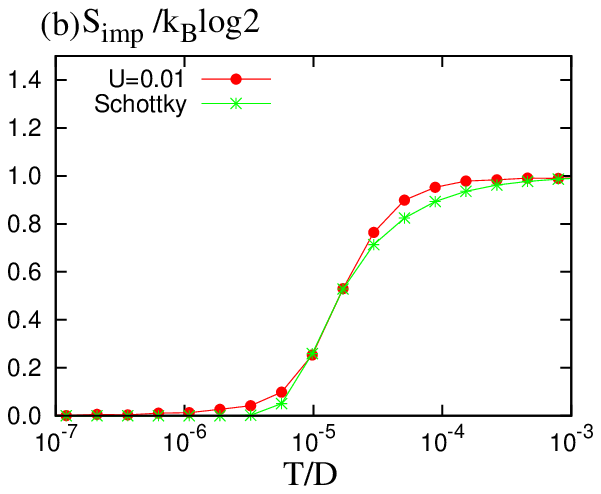}}
      \end{tabular}
   \end{center}
   \caption{(Color online) (a)Temperature dependence of entropy $S_{\text{imp}}$ for $U\leq1.2$ in the 
             strong electron-phonon coupling regime, $V_{0}=V_{1}=\omega=0.2$. Plateaus at $k_{\text{B}}\log2$ 
             originate from the degeneracy of the PD. At $U=0$ the binding energy of the PD is so large that 
             Yu and Anderson type Kondo temperature is lower than $10^{-10}$. For finite $U$, the degeneracy of 
             the PD is lifted by the Coulomb interaction. (b)Temperature dependence of $S_{\text{imp}}$ 
             for $U=0.01$ and its fit by a Shottoky type entropy with the excitation gap $\Delta=3.9\times10^{-5}$.
           }\label{entropy_lower_strong_U}
\end{figure}

\begin{figure}
   \begin{center}
      \begin{tabular}{cc}
         \resizebox{80mm}{!}{\includegraphics{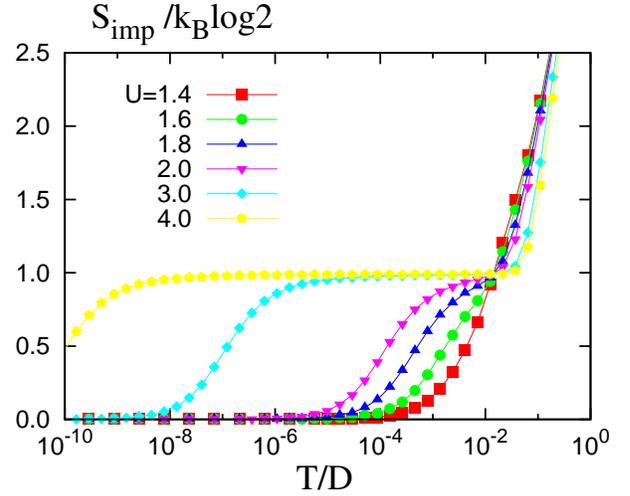}}
      \end{tabular}
   \end{center}
   \caption{(Color online) Temperature dependence of entropy $S_{\text{imp}}$ for $U\geq1.4$ in the strong 
             electron-phonon coupling regime, $V_{0}=V_{1}=\omega=0.2$. Plateaus at $k_{\text{B}}\log2$ come 
             from the degeneracy of the local moment for the relatively strong $U$. The low temperature energy 
             spectra are characterized by the $p$-type.}
   \label{entropy_upper_strong_U}
\end{figure}

In this subsection we show results of NRG calculations for a strong electron-phonon coupling case, 
$V_{0}=V_{1}=\omega=0.2$. In Figs. \ref{entropy_lower_strong_U}(a) and \ref{entropy_upper_strong_U} we see two distinct 
temperature dependence of $S_{\text{imp}}$ in the course of release of entropy from the plateau at 
$k_{\text{B}}\log2$.

For $U<1.4$ shown in Fig. \ref{entropy_lower_strong_U}(a), the electron-phonon coupling is relatively more important than 
the Coulomb interaction. In this region, the Coulomb interaction can be treated as a perturbation to the PD. 

For the non-interacting case the plateau of entropy at $k_{\text{B}}\log2$ is not suppressed 
even at low temperatures. Because of the strong electron-phonon coupling we expect that the YAK temperature is much lower 
than $10^{-10}$.

For $U>0$, the low temperature energy spectrum is of the $p$-type. In \S\ref{two-site}, we showed that the Coulomb interaction 
lifts the degeneracy of the PD in a similar way as a transversal magnetic field lifts degeneracy of a free spin. When $U$ 
is smaller than $1.4$, it is expected that low-energy physics of the impurity is identified as Yu and Anderson type Kondo effect 
under the "transverse field", which is characterized by $p$-YAK in the phase diagram, Fig. \ref{phase_diagram}. Therefore, 
it is useful for us to compare the present results with those of conventional spin $1/2$ Kondo model under magnetic field. 
It is shown for the latter model that when the magnetic field is larger than Kondo temperature behavior of heat capacity 
approaches to a Schottky form (see Fig. 15 in Ref. \ref{Bethe}). The lower panel of Fig. \ref{entropy_lower_strong_U} shows
$S_{\text{imp}}$ at $U=0.01$ fitted by the Schottky form with $\Delta=3.9\times10^{-5}$. Let us compare this $\Delta$ with 
the energy gap between the PD, namely $\Delta_{\text{PD}}$. To estimate $\Delta_{\text{PD}}$ it is appropriate to use 
a renormalized coupling constant $\lambda^{'}$ instead of the bare $\lambda(\equiv-V_{1}/\omega)$. We can estimate $\lambda^{'}$ 
from the NRG calculations by using the relation $\bigl<q^{2}\bigr>\equiv(16{\lambda^{'}}^{2}+1)/2$. In this way $\Delta_{\text{PD}}$
is estimated to be $2.62\times10^{-5}$ for $U=0.01$ by using $\bigl<q^{2}\bigr>=5.75$. It is important to note that there is a notable 
deviation at low temperatures from the Schottky form. It indicates that energy spectrum at low temperatures is not gapped but 
continuous, which is consistent with the fact that even with a finite $U(\neq0)$ the low temperature properties are described
by a local Fermi liquid coupled with the $p$-wave conduction electrons. 

We comment on the transition from the $s$-YAK regime to the $p$-YAK regime in the case of strong electron-phonon coupling.
Let us start from the noninteracting case, $U=0$. As we have discussed in \S\ref{non_int} Kondo temperature of the YAK effect 
$T_{\text{YAK}}$ is generally small and becomes extremely small as the binding energy $|E_{\text{PD}}|$ gets larger, see Fig. 
\ref{entropy_Yu_and_Anderson}(b). The Kondo energy, $k_{\text{B}}T_{\text{YAK}}$, is nothing but the energy gain at 
the $s$-type fixed points. On the other hand, the Coulomb interaction is favorable to the $p$-type fixed points and $\Delta_{\text{PD}}$
is its energy gain. Therefore, the transition takes place when $k_{\text{B}}T_{\text{YAK}}\sim\Delta_{\text{PD}}$. Since 
$k_{\text{B}}T_{\text{YAK}}$ is extremely small for strong electron-phonon coupling, the transition takes place close 
vicinity of the $U=0$ line.

For $U\geq1.5$ electrons in the impurity orbital behave as a local moment and it is quenched dominantly 
by the $p$-wave conduction electrons. Actually, in Fig. \ref{entropy_upper_strong_U} there are plateaus of entropy at
$k_{\text{B}}\log2$ and the entropy is released in the same temperature dependence as the usual Kondo effect.

\begin{figure}
   \begin{center}
      \begin{tabular}{cc}
         \resizebox{80mm}{!}{\includegraphics{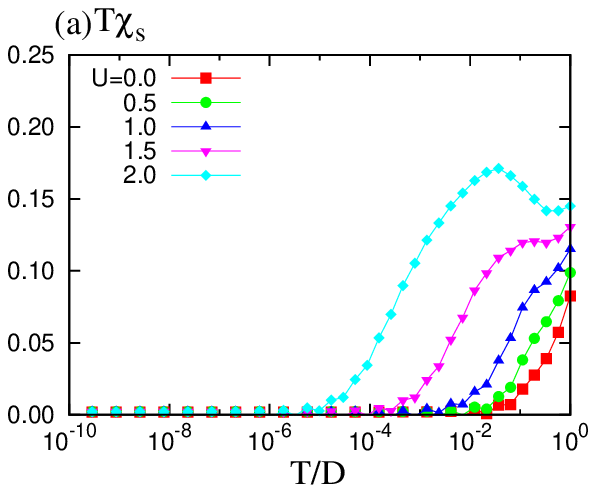}}\\
         \resizebox{80mm}{!}{\includegraphics{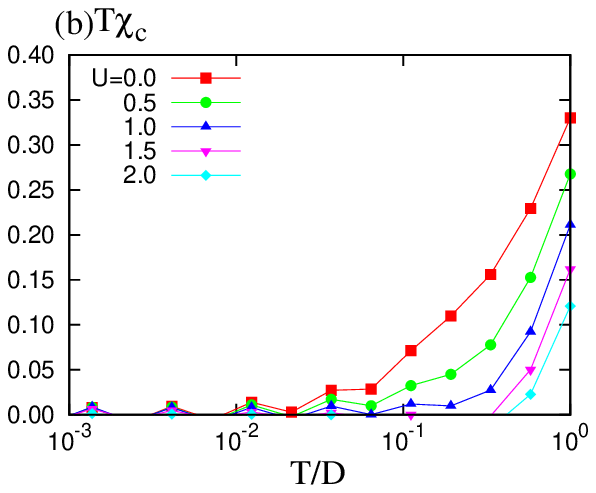}}
      \end{tabular}
   \end{center}
   \caption{(Color online) Temperature dependence of $T\chi_{s}$ (a) and $T\chi_{c}$ (b) in 
             the strong electron-phonon coupling regime, $V_{0}=V_{1}=\omega=0.2$. For the $U=0$, the relation
             $T\chi_{s}=0.25T\chi_{c}$ is valid. Unlike the weak electron-phonon coupling case, the enhancement of 
             $T\chi_{s}$ by Coulomb interaction is weak for $U<2.0$. $T\chi_{s}$ starts to show a maximum for $U>2.0$ 
             and becomes close to 0.25 as $U$ is increased.}\label{spin_charge_strong_U}
\end{figure}

Lastly, let us discuss the intermediate parameter region, see Fig. \ref{phase_diagram}. In the region between 
green and blue dashed lines, the entropy decays to zero without any plateau. It seems that the Coulomb interaction 
and the electron-phonon coupling are competing and almost balanced each other, leading to almost monotonic temperature 
dependence of $T\chi_{s}$ as shown in Fig. \ref{spin_charge_strong_U}(a) and $T\chi_{c}$ in (b). Of course 
the energy spectra at low temperatures are always characterized by the $p$-type except for $U=0.0$. 
However the nature of the physical properties changes from the PD with a small energy splitting due to 
the Coulomb interaction for $U\leq1.4$ to the local moment regime for $U\geq1.5$. The crossover between the two 
regimes is rapid but continuous.

\subsection{The line of 2-channel Kondo fixed points}
\begin{figure}
   \begin{center}
      \begin{tabular}{cc}
         \resizebox{80mm}{!}{\includegraphics{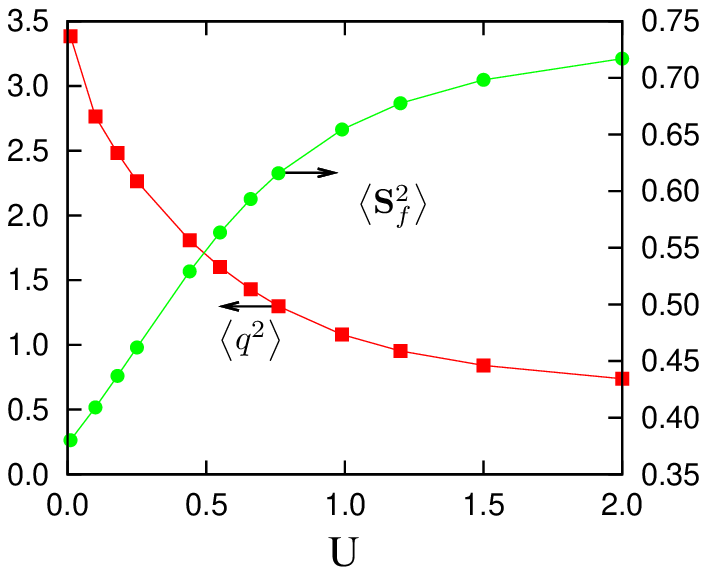}}\\
      \end{tabular}
   \end{center}
   \caption{(Color online) $U$ versus $\bigl<q^{2}\bigr>$ and $\bigl<\boldsymbol{S}^{2}_{f}\bigr>$ with $V_{0}=\omega=0.2$.
             The sets of the parameters, $V_{1}$ and $U$, are chosen to correspond to each points on the red 
             line with squares in Fig.\ref{phase_diagram}.}\label{local_quantity}
\end{figure}
Finally we discuss physical properties on the line of $2$-channel Kondo fixed points shown in Fig. \ref{phase_diagram}. 
On the line, energy spectra are always characterized by the $2$-chK type. For the strong Coulomb interaction regime, 
the effective Hamiltonian is given by eq. (\ref{Kondo}), which is equivalent to the conventional $2$-chK model. 
It is well known that the $2$-chK effect is realized when the two types of screening channels are fully balanced. 
In the weak correlation regime, however, the degrees of freedom screened by the conduction electrons are not 
the spin degrees of freedom of the impurity orbital but the ion displacements of the PD.

To observe the crossover behavior, we show thermal average of square of the ion displacement $\bigl<q^{2}\bigr>$ 
and magnitude of the local moment $\bigl<\boldsymbol{S}^{2}_{f}\bigr>$ calculated at low temperatures. Figure 
\ref{local_quantity} shows $\bigl<q^{2}\bigr>$ and $\bigl<\boldsymbol{S}^{2}_{f}\bigr>$ calculated for the sets of 
parameters on the line of $2$-chK fixed points in Fig. \ref{phase_diagram}. When $U$ increases along the line, 
$\bigl<q^{2}\bigr>$ is suppressed by Coulomb interaction and approaches to $0.5$ which corresponds to the zero point 
fluctuations. $\bigl<q^{2}\bigr>$ decreases monotonically, showing a concave curve as a function of $U$. 
On the other hand, the magnitude of the local moment $\bigl<\boldsymbol{S}^{2}_{f}\bigr>$ increases from 0.375 to $0.75$.
The former number corresponds to the free orbital limit, while the latter to the local moment limit. The curve of 
$\bigl<\boldsymbol{S}^{2}_{f}\bigr>$ is concave-downward. These results support the view that the degree of freedom to 
be screened by conduction electrons smoothly transfers from the ion displacement to the spin degrees of freedom of 
local moment with increasing $U$.

\section{Conclusion and future problems}\label{section_5}
In this paper we have discussed Kondo effects when a magnetic ion is vibrating in a harmonic potential. The central 
issue is the competition between the Coulomb and electron-phonon interactions. To treat the problem, 
a generalized impurity Anderson model is derived by taking account of vibrational mode of oscillations.
Associated with phonon absorption or emission, a new channel opens for hybridization of the localized orbital with 
conduction electrons. Suppose the localized orbital is described by an $\ell$-spherical wave, then the new channels 
are hybridization with the $\ell\pm1$ partial waves. In the simplest case of the $s$-wave symmetry of the localized orbital, 
the new channel is the mixing with the $p$-partial waves. 

To proceed further, we have restricted the ion oscillation to one dimension, in which case only one component among 
the three $p$-partial waves is relevant. With this reduction, the rotational symmetry of the original system is lowered 
to the inversion symmetry. This model still shows rich variety of interesting physics which includes different limiting 
cases: conventional $s$-wave Kondo effect, phonon-assisted $p$-wave Kondo effect and Yu-Anderson type Kondo effect.

The model has been analyzed by using the NRG scheme and it has been shown that the low-energy fixed points belong to 
one of the two different classes which are characterized by the $s$-type or $p$-type fixed points. On the boundary we find 
generically the $2$-chK fixed points. Surprisingly, not only the $s$-chK but also the YAK regions are continuously 
connected to the local Fermi liquid. It is also a new finding that the YAK effect which has been discussed for a system 
with strong electron-phonon coupling but without any effect of electron-electron correlation is actually very sensitive 
to the Coulomb interaction. With introduction of small $U$, the fixed point of the YAK effect changes from the $s$-type 
to the $p$-type. 

Concerning the Kondo effects of a magnetic ion vibrating in the sea of conduction electrons, there are a number of interesting 
questions to be addressed in future. Response to magnetic fields may be different depending on where the system is 
located in the phase diagram. Especially, effect of magnetic field for the YAK effect would be very interesting in the 
region around the boundary of the 2-chK. For the present model interesting phase transitions take place close to the 
region of intermediate electron-phonon coupling, $V_{0}\sim V_{1}$ and $V_{1}\sim\omega$. When we consider a strongly 
anharmonic potential like a double well one, the condition for the occurrence of the YAK effect would be drastically 
modified. In view of the fact that there are some experimental indications that magnetic ions in some skutterudite compounds 
migrate among off-center positions, this might be an interesting question relevant to this class of 
materials\cite{off_center1,off_center2,off_center3}.

\section{Acknowledgements}
The authors would like to thank C. Hori for many helpful discussions. This work is supported by Grant-in-Aid on Innovative Areas 
"Heavy Electrons" (No.$20100208$) and also by Scientific Research (C) (No.$20540347$). S.Y. acknowledges support from Global 
COE Program "the Physical Sciences Frontier", MEXT. S.K. is supported by JSPS Grant-in-Aid for JSPS Fellows 
$21\cdot6752$ and K.H. by KAKENHI ($20740189$).

\appendix
\section{Expansion of the overlap integrals within $O(\boldsymbol{Q})$}
In this appendix we derive hybridization terms expanded within the first order of ion displacement by using 
the generalized gradient theorem\cite{Bayman}. This theorem is applicable to the present model because the rotational 
symmetry $SO(3)$ is assumed in the total system including both the ion and the conduction electrons.
First, let us transform the coordinate system from the conventional Cartesian form to the spherical tensor form. 
The position vectors of a conduction electron and the ion displacement $\boldsymbol{Q}$ are 
transformed to
\begin{align}
r_{1,\pm 1} &= \mp\frac{1}{\sqrt{2}}(r_{x}\mp \text{i}r_{y}),\ \ \ r_{1, 0} = r_{z},\notag\\
Q_{1,\pm 1} &= \mp\frac{1}{\sqrt{2}}(Q_{x}\pm \text{i}Q_{y}),\ \ \ Q_{1, 0} = Q_{z}.\notag
\end{align} 
One can show that the set of differential operators $\{\nabla_{1,m}\}$ defined by $\{r_{1,m}\}$ 
are first rank spherical tensor operators. Within $O(\boldsymbol{Q})$ the overlap integral is written as
\begin{align}
 & V_{(\boldsymbol{k};\ell m)}(\boldsymbol{Q})\notag\\
=& \frac{g}{\sqrt{\Omega}}\int d^{3}\boldsymbol{r}\text{e}^{-\text{i}\boldsymbol{k}\cdot\boldsymbol{r}}
             \psi_{\ell m}(\boldsymbol{r}-\boldsymbol{Q})\notag\\
\simeq 
 & \frac{g}{\sqrt{\Omega}}\int d^{3}\boldsymbol{r}\text{e}^{-\text{i}\boldsymbol{k}\cdot\boldsymbol{r}}
             \bigl\{\psi_{\ell m}(\boldsymbol{r})-\boldsymbol{Q}\cdot\nabla\psi_{\ell m}(\boldsymbol{r})\bigr\}\notag.
\end{align}
$\boldsymbol{Q}\cdot\nabla$ is expressed in the spherical tensor form by $\sum_{m}Q^{*}_{1,m}\nabla_{1,m}$
because of the $SO(3)$ symmetry. 

The spherical tensors can be expressed in terms of the spherical harmonics $|\ell,m\bigr>$ and
the Clebsch-Gordan coefficients $\bigl<\ell_{1},m_{1};\ell_{2},m_{2}|\ell,m\bigr>$ by
\begin{align}
 &\bigl<\ell',m'|\nabla_{1,m_{1}}|\ell,m\bigr>\notag\\
=&\sqrt{\frac{\ell+1}{2\ell+3}}\bigl<\ell,m;1,m_{1}|\ell',m'\bigr>
      \biggl\{\frac{\partial}{\partial r}  - \frac{\ell}{r}\biggr\}\notag\\
 &\ \ \ \ \ \ \ \ \ \ \ \ \ \ \ \ \times\delta_{\ell',\ell+1}\delta_{m',m+m_{1}}\notag\\
 &-\sqrt{\frac{\ell}{2\ell-1}}\bigl<\ell,m;1,m_{1}|\ell',m'\bigr>
      \biggl\{\frac{\partial}{\partial r}  + \frac{\ell+1}{r}\biggr\}\notag\\
 &\ \ \ \ \ \ \ \ \ \ \ \ \ \ \ \ \times\delta_{\ell',\ell-1}\delta_{m',m+m_{1}}.
\end{align}
After rewriting $\text{e}^{-\text{i}\boldsymbol{k}\cdot\boldsymbol{r}}$  by using the spherical wave expansion, 
the integrals expanded within the first order of $\boldsymbol{Q}$ are given by
\begin{align}
 &V_{(\boldsymbol{k};\ell m)}(\boldsymbol{Q})\notag\\
=&\text{i}^{-\ell}R_{\ell}(k)\sqrt{4\pi}Y_{\ell m}(\hat{\boldsymbol{k}})
   +\sum^{1}_{m_{1}=-1}Q^{*}_{1,m_{1}}\notag\\
 &\times\biggl\{\varGamma^{+}_{(\ell,m;m_{1})}
               \text{i}^{-(\ell +1)}r_{\ell +1}(k)\sqrt{4\pi}Y_{\ell+1,m+m_{1}}(\hat{\boldsymbol{k}})\notag\\
 &+\varGamma^{-}_{(\ell,m;m_{1})}
               \text{i}^{-(\ell-1)}r_{\ell-1}(k)\sqrt{4\pi}Y_{\ell-1,m+m_{1}}(\hat{\boldsymbol{k}})  
               \biggr\},
\end{align}
where $\varGamma^{+(-)}_{(\ell,m;m_{1})}$ are defined by
\begin{align}
\varGamma^{+}_{(\ell,m;m_{1})} &= \sqrt{\frac{\ell+1}{2\ell+3}}\bigl<\ell,m;1,m_{1}|\ell+1,m+m_{1}\bigr>,\\
\varGamma^{-}_{(\ell,m;m_{1})} &= \sqrt{\frac{\ell  }{2\ell-1}}\bigl<\ell,m;1,m_{1}|\ell-1,m+m_{1}\bigr>.
\end{align}
The radial integrals are represented by the spherical Bessel functions and the radial parts of 
$\psi_{\ell m}(\boldsymbol{r})$ as
\begin{align}
&R_{\ell  }(k) \equiv \sqrt{\frac{4\pi}{V}}g  \int drr^{2}j_{\ell}(kr)u_{\ell}(r),\\
&r_{\ell+1}(k)\notag\\ \equiv
&\sqrt{\frac{4\pi}{V}}g  \int drr^{2}j_{\ell+1}(kr)
            \biggl\{\frac{\ell}{r} - \frac{\partial}{\partial r}  \biggr\}u_{\ell}(r),\\
&r_{\ell-1}(k)\notag\\ \equiv 
&\sqrt{\frac{4\pi}{V}}g  \int drr^{2}j_{\ell-1}(kr)
            \biggl\{\frac{\partial}{\partial r}  + \frac{\ell+1}{r}\biggr\}u_{\ell}(r).
\end{align}
Lastly we neglect the $k$ dependence of the radial integrals and use the values at the Fermi 
momentum $k_{\text{F}}$. After introducing the dimensionless ion displacement $\boldsymbol{q}(\equiv \alpha\boldsymbol{Q})$ 
we obtain eq. (\ref{hybridization}), where $V_{\ell}$ and $V^{(m;m')}_{\ell\pm1}$ are defined by
\begin{align}
        V        _{\ell} 
&\equiv \text{i}^{-\ell}R_{\ell     }(k_{\text{F}}),\\
        V^{(m;m')}_{\ell\pm 1} 
&\equiv \text{i}^{-(\ell\pm 1)}\frac{r_{\ell\pm 1}(k_{\text{F}})}{\alpha}\varGamma^{\pm}_{(\ell,m;m_{1})}\label{V1}.
\end{align}

\end{document}